\newcommand{\be}{\begin{equation}}
\newcommand{\ee}{\end{equation}}
\newcommand{\Mo}{\ {\rm M_\odot}}
\def\LCDM{$\Lambda$CDM}
\def\kpc{\ {\rm kpc}} \def\pc{\ {\rm pc}}
\newcommand{\kms}{\>{\rm km}\,{\rm s}^{-1}}
\documentclass{mn2e}
\input{epsf}
\title[Mass modelling of dSph galaxies]
{Mass modelling of dwarf spheroidal galaxies: the effect of unbound stars from tidal tails and
the Milky Way}

\author[J. Klimentowski et al.]
    {Jaros{\l}aw Klimentowski,$^{1}$ Ewa L. {\L}okas,$^{1}$ Stelios Kazantzidis,$^{2}$ Francisco
    Prada,$^{3}$ \newauthor{Lucio Mayer}$^{4,5}$ and Gary A. Mamon$^{6,7}$
    \\
    \\
    $^1$Nicolaus Copernicus Astronomical Center, Bartycka 18,
    00-716 Warsaw, Poland\\
    $^2$Kavli Institute for Particle Astrophysics and Cosmology, Department of Physics,
	Stanford University, P.O. Box 20450, M/S 29,\\ Stanford, CA 94309, USA\\
    $^3$Instituto de Astrof{\'\i}sica de Andalucia (CSIC),
    Apartado Correos 3005, E-18080 Granada, Spain\\
    $^4$Institute for Theoretical Physics, University of Z\"urich, CH-8057 Z\"urich, Switzerland\\
    $^5$Institute of Astronomy, Department of Physics, ETH Z\"urich, Wolfgang-Pauli
    Strasse, CH-8093 Z\"urich, Switzerland  \\
    $^6$Institut d'Astrophysique de Paris (UMR 7095: CNRS and Universit\'e Pierre \& Marie Curie),
    98 bis Bd Arago,
    F-75014 Paris, France \\
    $^7$GEPI (UMR 8111: CNRS and Universit\'e Denis Diderot), Observatoire de Paris,
    F-92195 Meudon, France}
\begin{document}

\maketitle

\begin{abstract}
We study the origin and properties of the population of unbound stars in the kinematic samples
of dwarf spheroidal galaxies. For this purpose we have run a high resolution $N$-body simulation
of a two-component dwarf galaxy orbiting in a Milky Way potential. In agreement with the tidal
stirring scenario of Mayer et al., the dwarf is placed on a highly eccentric orbit, its initial
stellar component is in the form of an exponential disk and it has a NFW-like dark matter halo.
After 10 Gyrs of evolution the dwarf produces a spheroidal stellar component and
is strongly tidally stripped so that mass follows light and the stars are on almost isotropic
orbits. From this final state, we create mock kinematic data sets for 200 stars by observing the
dwarf in different directions. We find that when the dwarf is observed along the tidal tails the
kinematic samples are strongly contaminated by unbound stars from the tails. We also study another
source of possible contamination by adding stars from the Milky Way. We demonstrate
that most of the unbound stars can be removed by the method of interloper rejection proposed by
den Hartog \& Katgert and recently tested on simulated dark matter haloes. We model the cleaned-up
kinematic samples using solutions of the Jeans equation with constant mass-to-light ratio
and velocity anisotropy parameter. We show that even for such a strongly stripped dwarf the Jeans analysis,
when applied to cleaned samples, allows us to reproduce the mass and
mass-to-light ratio of the dwarf with accuracy typically better than 25 percent and almost exactly
in the case when the line of sight is perpendicular to the tidal tails.
The analysis was applied to the new data for the Fornax dSph galaxy. We show that
after careful removal of interlopers the velocity dispersion profile of Fornax can be
reproduced by a model in which mass traces light with a mass-to-light
ratio of 11 solar units and isotropic orbits. We demonstrate that most of the contamination
in the kinematic sample of Fornax probably originates from the Milky Way.

\end{abstract}


\section{Introduction}

Dwarf spheroidal (dSph) galaxies of the Local Group provide critical tests of theories of
structure formation in the Universe. The number density of dwarfs in the vicinity of the
Milky Way (MW) poses a problem for theories based on Cold Dark Matter
(Klypin et al. 1999; Moore et al. 1999). In particular, $\Lambda$CDM $N$-body simulations predict
a few hundred dwarf galaxies in the Local Group while only a few tens are observed.
Their numbers, distribution, internal dynamics and density profiles have important
implications also for other issues like gravitational lensing and dark matter (DM) detection
experiments.

A number of solutions to the problem have been proposed (for a recent review see Kravtsov, Gnedin \&
Klypin 2004), but since the degree of discrepancy with the models depends on precise knowledge
of dwarf galaxy masses one of the important issues is to accurately determine the masses.
It is now generally believed that Galactic dSph galaxies are dominated by
DM, but it is still debated to what extent their large velocity dispersions
are due to their high DM content and how much they can be contaminated by
the presence of unbound stars.

The unbound stars may originate from the dwarf itself due to
tidal stripping in the potential of the giant host galaxy around which the dwarf orbits
or from the stellar populations of the host galaxy. The latter source of contamination can
be dealt with by careful photometric analysis. Recent studies have shown that restricting the
target stars for spectroscopic observations by the colour-magnitude diagram of the dwarf galaxy
does not solve the problem (Kleyna et al. 2002) but additional colour-colour diagrams can help
distinguish between the dwarf galaxy giant star population and the MW dwarf stars
(Majewski et al. 2000a; Mu\~noz et al. 2006).

The contamination due to tidally stripped stars however remains poorly understood although convincing
evidence for the presence of tidal tails exists for some dSph galaxies. Majewski et al. (2000b) and
Mu\~noz et al. (2006) find a significant excess of Carina member stars above the nominal King radius
while Coleman et al. (2005) find excess number density of stars around Fornax depending on the direction
of the measurements. Mart{\'\i}nez-Delgado et al. (2001) detected the tidal extension in the Ursa Minor
dwarf and G\'omez-Flechoso \& Mart{\'\i}nez-Delgado (2003) modelled the properties of tidal tails in this
galaxy to estimate its mass-to-light ratio. Detection of tidal tails remains extremely hard though
due to very low surface brightness of the tails and critical dependence on the assumed background
level which is very difficult to estimate reliably.

Modelling of the dSph kinematics by the Jeans analysis is still the major tool to determine
their mass content. These are usually performed on stellar samples selected by a simple
cut-off in velocity
(with respect to the mean systemic velocity of the dwarf) or on stars with velocities within 3
standard deviations from the mean (e.g. Wilkinson et al. 2004). In the recent study of the Draco
dSph {\L}okas, Mamon \& Prada (2005) using the sample selected by Wilkinson et al.
and modelling both the dispersion and kurtosis of the line-of-sight
velocity distribution showed that no realistic, consistent solution to the Jeans equations can be found
unless some stars are rejected from the sample. Along the same lines, Mu\~noz et al. (2006) argued
that for all of the detected Carina stars to be bound the dwarf would have to possess an enormous
mass-to-light ratio of a few thousand solar units. These arguments strongly suggests the presence
of unbound stars in the traditionally selected samples.

The purpose of this work was to study the origin of unbound stars and test the methods of dealing
with them using $N$-body simulations. The paper is organized as follows.
In section~2 we describe the simulation details. Section~3
presents the main properties of the simulated dwarf at its final stage.
In section~4 we describe the creation of the mock kinematic data and apply a method for
interloper removal to data sets contaminated by the tidally stripped stars and the stellar
populations of the MW. Results
of fitting the models to the cleaned data are presented in section~5. In section~6
we apply the developed procedure to the real data for the Fornax dSph galaxy.
The discussion follows in section~7.

\section{The simulation}

The `tidal stirring' model originally proposed by Mayer et al. (2001)
constitutes a viable mechanism for producing the majority of dSphs
in the Local Group. These authors performed $N$-body simulations of the dynamical evolution
of dwarf galaxies in a MW potential and concluded that the dSph progenitors were
rotationally supported low surface brightness dwarf systems that were accreted by the DM halo of the
dominant spiral galaxy at early times. Strong gravitational interactions with the primary host potential
at pericentre heat the stellar disks and produce objects whose stellar structure and kinematics resemble
those of dSph galaxies. Very recently, variants of this model have been shown to successfully explain
properties such as the morphology-density relation and the extreme mass-to-light ratios inferred
for some of the dwarfs (Mayer et al. 2006, 2007).

In this investigation, we adopt the tidal stirring model of Mayer et al. (2001)
and perform a high-resolution $N$-body simulation of a dwarf disk-like system
orbiting within the static host potential of a MW-sized galaxy. Our approach
thus neglects the effects of dynamical friction and the response of the primary to
the presence of the satellite. However, this choice is justified because orbital decay
times are expected to be longer than the Hubble time given  the difference in mass
between the two systems and the additional mass loss due to tidal stripping (Mayer et
al. 2001; Hayashi et al. 2003; Kazantzidis et al. 2004).

Live dwarf galaxy models are constructed using the technique by Hernquist (1993) and consist
of an exponential stellar disk embedded in a spherical and isotropic Navarro et al. (1997, hereafter NFW)
DM halo. The structural properties of dark haloes and disks are motivated by the
currently favoured concordance {\LCDM} ($\Omega_{\rm M}=0.3$, $\Omega_{\Lambda}=0.7$,
$h=0.7$) cosmological model (Mo, Mao \& White 1998). The density distribution of the NFW profile
is given by
\begin{equation}    \label{equation:NFW_profile}
	\rho(r)=\frac{\rho_{\rm s}} {(r/r_{\rm s}) (1+r/r_{\rm s})^2} ,
\end{equation}
where $\rho_{\rm s}$ is a characteristic inner density
and $r_{\rm s}$ denotes the scale radius of the system. The NFW density law has a
cumulative mass profile that diverges at large radii. In order to keep the total mass
finite, we model the halo density profile beyond the virial radius of the system,
$r_{\rm vir}$, using an exponential cutoff (Kazantzidis et al. 2004). The concentration
parameter $c = r_{\rm vir}/r_{\rm s}$ controls the shape of the halo density
profile.

The stellar disk follows an exponential distribution in cylindrical radius $R$
and its vertical structure is modelled by isothermal sheets
\begin{equation}    \label{disk_density}
	\rho_{\rm d}(R,z) = \frac{M_{\rm d}}{4\pi R_{\rm d}^2 z_{\rm d} }
	\exp\left(-\frac{R}{R_{\rm d}}\right) \ {\rm sech}^2\left(\frac{z}{z_{\rm d}}\right),
\end{equation}
where $M_{\rm d}$, $R_{\rm d}$ and $z_{\rm d}$ denote the mass, radial scale length, and
vertical scale height of the disk, respectively. In our modelling, we parametrize the disk
mass to be a fraction $m_{\rm d}$ of the halo virial mass, $M_{\rm d}=m_{\rm d}M_{\rm
vir}$, and we specify the disk vertical scale height in units of the radial disk scale
length.

Once a set of cosmological parameters is adopted, the virial quantities $M_{\rm
vir}$ and $r_{\rm vir}$ are uniquely determined by the halo circular velocity at the
virial radius, $V_{\rm vir}$ (Mo et al. 1998). Furthermore, the DM halo carries some net
angular momentum specified by the dimensionless spin parameter, $\lambda = J/G \sqrt{|E|/
M_{\rm vir}^5}$, where $J$ and $E$ are the total halo angular momentum and energy,
respectively. We follow Springel \& White (1999) and distribute the angular momentum of
the DM halo by setting the halo streaming velocity to be a fixed fraction of the local
total circular velocity. Our models implicitly assume that the disk forms out of collapsed gas which
started with the same specific angular momentum as the halo and that such angular
momentum was conserved during infall (Mo et al. 1998).
Finally, the DM halo is adiabatically
contracted in response to  the growth of the stellar disk under the assumptions of
spherical symmetry, homologous contraction, circular DM  particle orbits, and angular
momentum conservation (Blumenthal et al. 1986). Adiabatic contraction allows to
construct {\it nearly} self-consistent disk galaxy models (see Kazantzidis et al. 2006
for a complete discussion). For further details on the adopted modelling we refer the reader to
Hernquist (1993) and Kazantzidis et al. (2006).

The dwarf galaxy halo
has a virial velocity of $V_{\rm vir}=20\kms$, implying a virial mass  and radius of
$M_{\rm vir} \simeq 3.7 \times 10^{9} \Mo$ and $r_{\rm vir} \simeq 40.2 \kpc$
respectively, and a concentration of $c=15$, resulting in a scale radius of $r_{\rm s}
\simeq 2.7 \kpc$ and a maximum  circular velocity of $V_{\rm peak} \simeq 30 \kms$. The
latter value is in the range of those inferred for some of the progenitors of dSphs from
high resolution $N$-body simulations (Kazantzidis et al. 2004;  Kravtsov et al. 2004).
After entering the host potential, $V_{\rm peak}$ will decrease due to mass loss
processes. Moreover, once the scatter in halo concentrations reflecting the different
formation epochs is considered, the adopted value of $c$ is consistent with the typical values
expected at these mass scales at $z \sim 2$ (Col\'{\i}n et al. 2004). We fix the disk
mass fraction to $m_{\rm d}=0.04$ which is typical of mass  models that reproduce dwarf
and low surface brightness galaxies (Jimenez, Verde \& Oh 2003). The exponential disk scale
length  is then fixed by the adopted spin parameter, $\lambda=0.08$, which is chosen to
be larger than the average value of  halo spins of $\sim 0.045$ found in cosmological
$N$-body simulations (e.g. Vitvitska et al. 2002) for two reasons. First, modelling of
rotation curves of dwarf galaxies suggests that they have an average spin larger than the
mean value  of the galaxy population as a whole (Jimenez et al. 2003). Second, using a
large spin parameter enables us  to construct a low surface brightness model as required
by the tidal stirring scenario. Finally, we adopt a constant vertical scale height across
the disk equal to $z_{\rm d}=0.1 R_{\rm d}$.  The radial disk scale length $R_{\rm d}$ is
uniquely determined for a given set of parameters $M_{\rm vir}$, $c$, $\lambda$, and
$m_{\rm d}$ and is equal to $R_{\rm d} \sim 1.3 \kpc$. The setup of the stellar disk is
complete once the Toomre  parameter, $Q$, is set (Toomre 1964). The mean Toomre parameter
of the disk was about $1.5$.

The external, spherically symmetric static tidal field is based on the dynamical mass model
A1 for the MW presented in Klypin, Zhao \& Somerville (2002).
Specifically, the DM halo has  a virial mass of $M_{\rm vir}=10^{12}\Mo$, a concentration
parameter of $c=12$, and a dimensionless spin parameter of $\lambda=0.031$. The halo was
adiabatically contracted to respond to the growth of the stellar disk and bulge. The mass
and thickness of the stellar disk were $M_{\rm d}=0.04 M_{\rm vir}$ and $z_{\rm d}=0.1
R_{\rm d}$, respectively, and $R_{\rm d}=3.5\kpc$ was the resulting disk scale length.
The mass and scale radius of the bulge are specified by $M_{\rm b}=0.008 M_{\rm vir}$ and
$a_{\rm b} = 0.2 R_{\rm d}$.

The orbits of the MW dSphs are currently poorly
constrained observationally. Nevertheless, their current distances, which give an
indication of the apocentre of their orbits, coupled with studies of the orbital
properties of cosmological haloes, can be used to constrain the orbital parameters of
dSphs. Using this information we placed the dwarf galaxy on an orbit bound and coplanar
with respect to the MW disk with the apocentre radius of $r_{\rm apo}=120\kpc$ and $r_{\rm
apo}/r_{\rm per}=\, 5/1$, close to the median ratio of apocentric to pericentric radii
found in high resolution cosmological $N$-body simulations (Ghigna et al. 1998). We begin
the simulation by randomly orienting the disk  of the satellite and placing it at
apocentre, and we integrate the orbit forward for $10$ Gyr.  This timescale corresponds to
about five orbital periods ($T_{\rm orb} \sim 2$ Gyr) in the chosen orbit and
represents a significant fraction of cosmic time.

The simulation was performed
using PKDGRAV, a multi-stepping, parallel, tree $N$-body  code (Stadel 2001). We sampled
the dwarf galaxy with $N = 4 \times 10^6$ DM particles and $N = 10^6$ stellar particles,
and employed a gravitational softening length of 100 and 50 pc,  respectively.
Numerical and structural parameters of the static primary and live dwarf galaxy model are
summarized in Table~\ref{galaxypar}.

\begin{table}
\begin{center}
\caption{Parameters of galaxy models.}
\begin{tabular}{lll}
System  &  Parameter & Value  \\
\hline
Dwarf: 	        &  $V_{\rm vir}$  & $20\kms$ \\
		&  $c$            & $ 15$    \\
		& $\lambda$       & $0.031$  \\
		&  $m_{\rm d}$    & $0.04$   \\
		&  $M_{\rm vir}$  & $ 3.7\times 10^{9}\Mo$ \\
		&  $r_{\rm vir}$  & $ 40.2\kpc$ \\
		&  $r_{\rm s}$    & $ 2.7 \kpc$ \\
		&  $V_{\rm peak}$ & $ 30\kms$ \\
		&  $R_{\rm d}$    & $ 1.3\kpc$ \\
		&  $z_{\rm d}$    & $ 0.13\kpc$ \\
		&  $Q$            & $ 1.5$ \\
		& $N_{\rm DM}$    & $4\times10^{6}$ \\
		&  $\epsilon_{\rm DM}$  & $100$ pc \\
		&  $N_{\rm d}$    & $10^{6}$ \\
		&  $\epsilon_{\rm d}$ & $50\pc$\\
\hline
Milky Way:      &  $M_{\rm vir}$  & $ 10^{12}\Mo$ \\
		&  $c$            & $ 12$ \\
		&  $\lambda$      & $ 0.031$ \\
		&  $M_{\rm d}$    & $ 4\times 10^{10}\Mo$  \\
		&  $R_{\rm d}$    & $ 3.5\kpc$ \\
		&  $z_{\rm d}$    & $ 0.35\kpc$ \\
		&  $M_{\rm b}$    & $ 8\times 10^{9}\Mo$  \\
		&  $a_{\rm b}$    & $700\pc$ \\
\hline
\label{galaxypar}
\end{tabular}
\end{center}
\end{table}

\begin{figure}
    \leavevmode
    \epsfxsize=7.2cm
    \epsfbox[0 0 340 370]{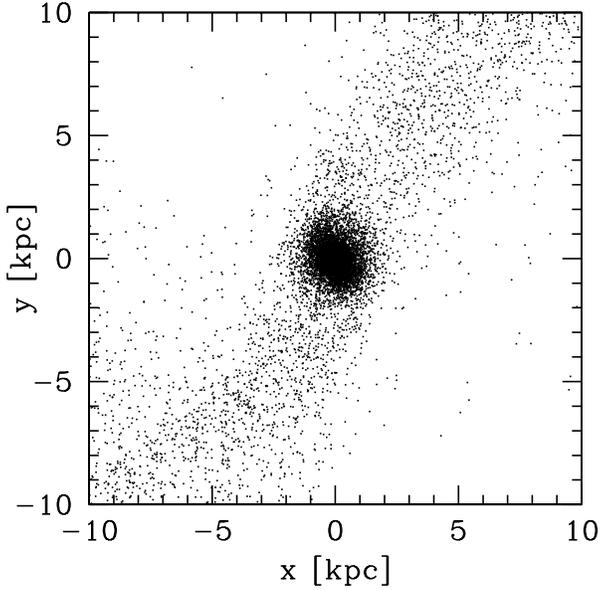}
\caption{Close-up of the stellar component of the simulated dSph galaxy.
Only every tenth star was plotted.}
\label{dwarfplot}
\end{figure}

\section{Properties of the simulated dwarf}

In this section we summarize the properties of the dSph galaxy formed in the simulation.
Fig.~\ref{dwarfplot} shows the stellar component of the dwarf galaxy and its neighbourhood
in the final output of the
simulation. The spheroidal component and the tidal tails are clearly visible. We start
by measuring the density distribution of stars and DM in the remnant and comparing them
with the initial distributions. The centre of the mass distribution was determined iteratively
by calculating the centre of mass of particles enclosed in a sphere of a given size, choosing a
new smaller sphere around thus determined centre and repeating the procedure until reasonable
convergence is reached. The velocities used in the following analysis were calculated with respect to
the mean velocity of particles within 0.5 kpc from the dwarf centre. All the quantities were plotted
starting from $r=200$ pc, i.e. twice the softening scale for DM particles so that they are
not affected by resolution.

We consider both the distributions
of all particles found in the vicinity of the dwarf and of particles bound to the dwarf
that can be actually counted as belonging to it.
In order to determine which particles are bound we treat the dwarf as an isolated object and
calculate the escape velocities in the standard way.
The potential energy of a given particle is calculated by summing up the contribution from
all the other particles in a large box of 20 kpc on a side. This is a straightforward,
although computationally expensive, procedure which needs to be applied in the case of a non-spherical
object. The particles identified as unbound are then
removed from the sample and the calculation is repeated until no more particles are removed.

\begin{figure}
\begin{center}
    \leavevmode
    \epsfxsize=6.6cm
    \epsfbox{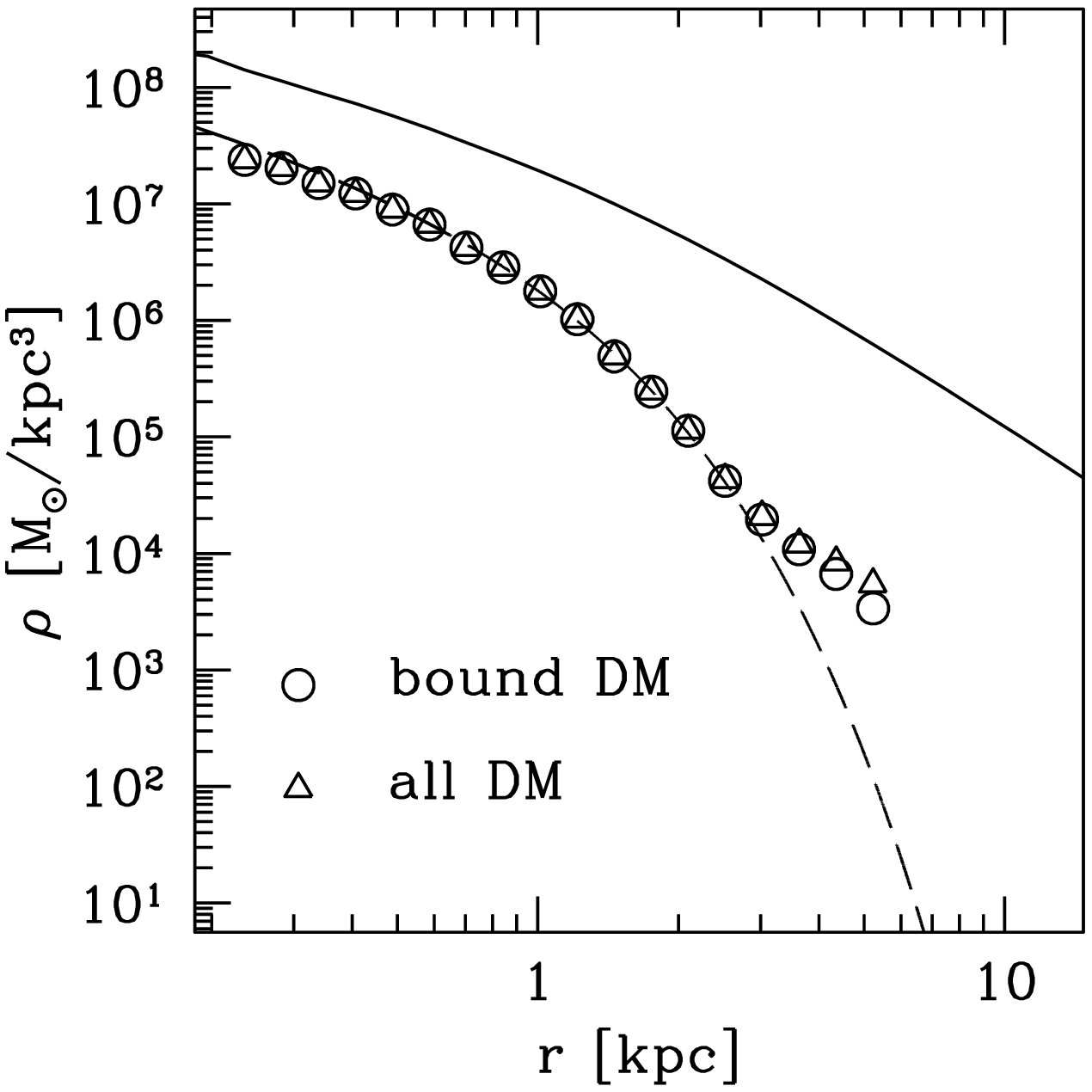}
    \epsfxsize=6.6cm    \epsfbox{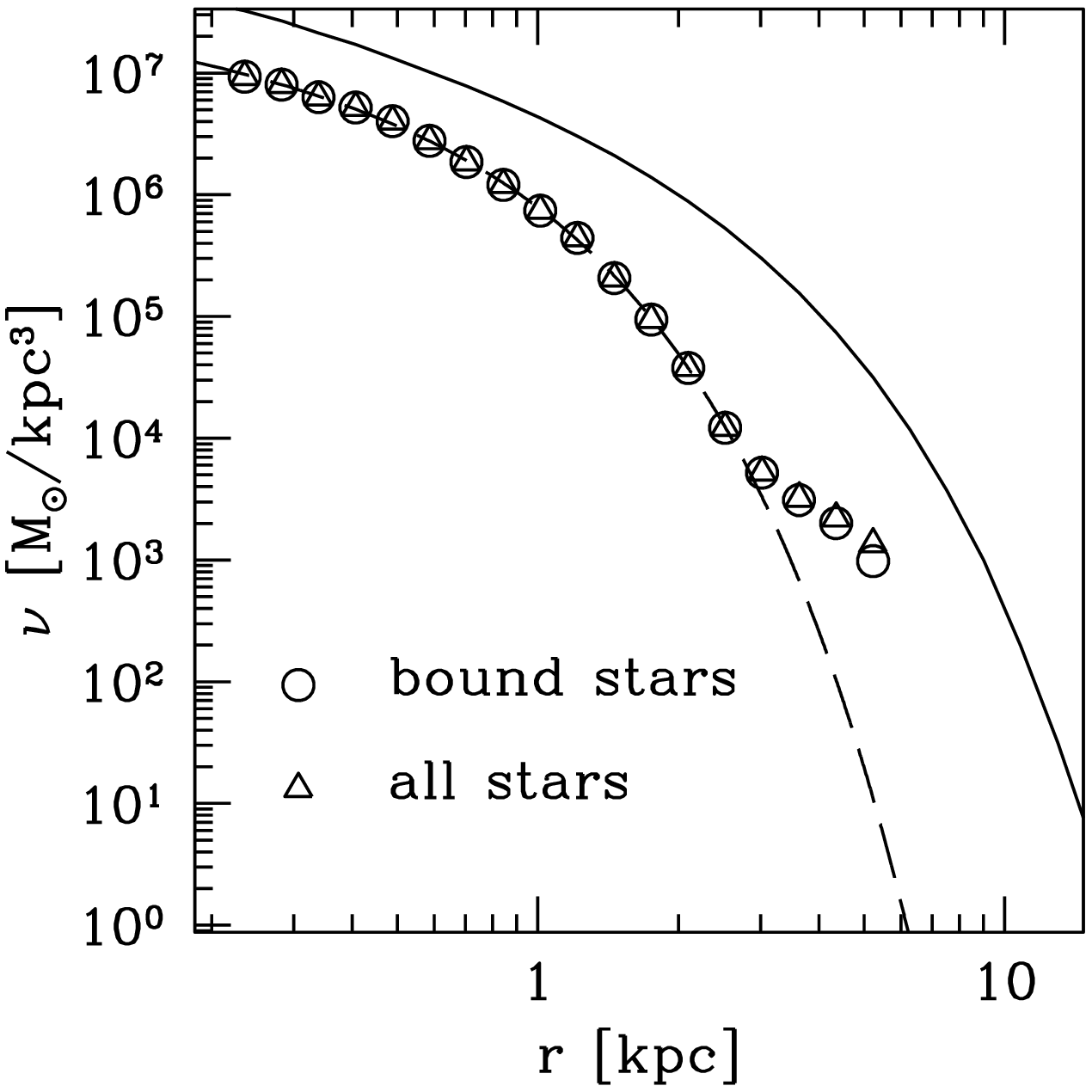}
    \epsfxsize=6.6cm
    \epsfbox{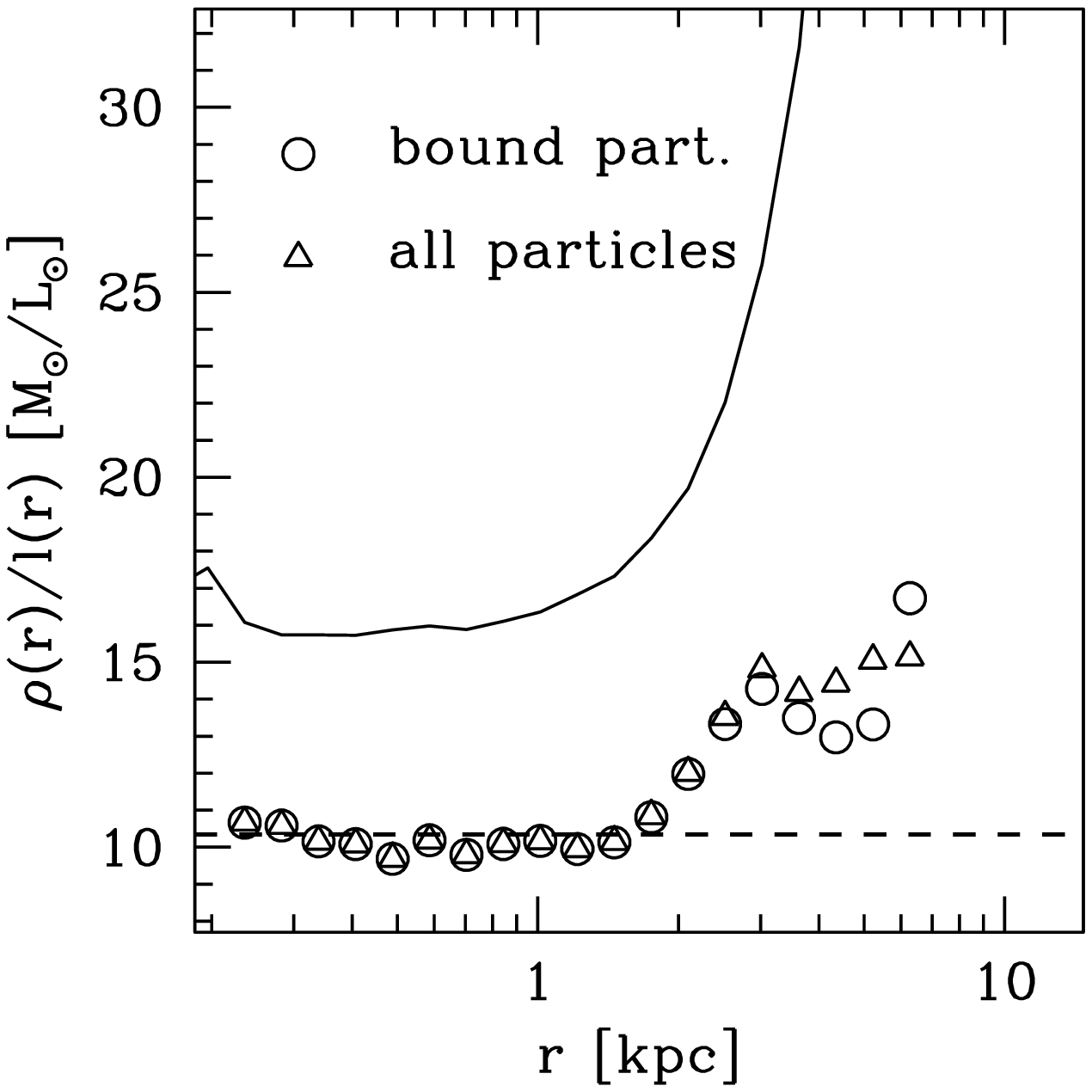}
\end{center}
\caption{Upper panel: the density distribution of DM particles in the final stage of the
dwarf. Open circles show the results for bound particles, open triangles for all dark particles
in the vicinity of the dwarf. The dashed line is the best fit with formula (\ref{kazantzidis}).
Middle panel: the same as the upper panel, but for stellar particles. The dashed line is the best
fit obtained with the S\'ersic profile (\ref{depsersic}). Lower panel: mass-to-light density
ratios measured for the bound (open circles) and for all particles (triangles). The dashed line
shows the best-fitting constant value for $r<2.5$ kpc. In each panel the solid line plots the initial
profile.}
\label{density}
\end{figure}

The distribution of bound DM particles is shown in the upper panel of Fig.~\ref{density}
with open circles. For comparison we also plot the distribution of all DM particles found in
the vicinity of the dwarf with open triangles. We can see that both distributions preserve the
cusp of the initial distribution at small radii: fitting a few inner data points
with $\rho \propto r^\alpha$
we obtain $\alpha \approx -1.2$ both for the initial and final DM distribution in the dwarf in
agreement with Kazantzidis et al. (2004) who found that the cusp is robust against tidal disruption.
At larger radii the final DM profiles steepen strongly and flatten at around
$r=2-3$ kpc due to the presence of tidal tails. Although the effect is weaker for the bound
particles, it is clearly visible, i.e. some of the particles in the tails are bound to the dwarf.
Rejecting the last three data points for the bound particles we fitted to the distribution an
analytic formula proposed by Kazantzidis et al. (2004)
\begin{equation}    \label{kazantzidis}
    \rho_{\rm d}(r) = C r^{-1} {\rm exp} \left( -\frac{r}{r_{\rm b}} \right),
\end{equation}
which was found to fit a density distribution of a stripped DM halo in a similar simulation
but with only DM.
The dashed line shows the fitted density profile with parameters $C=1.2 \times 10^7$
M$_{\sun}/$kpc$^2$ and $r_{\rm b} = 0.53$ kpc.
For comparison we also show with a solid line the initial NFW-like density profile of the progenitor.

\begin{figure}
\begin{center}
    \leavevmode
    \epsfxsize=7.5cm
    \epsfbox[0 0 370 370]{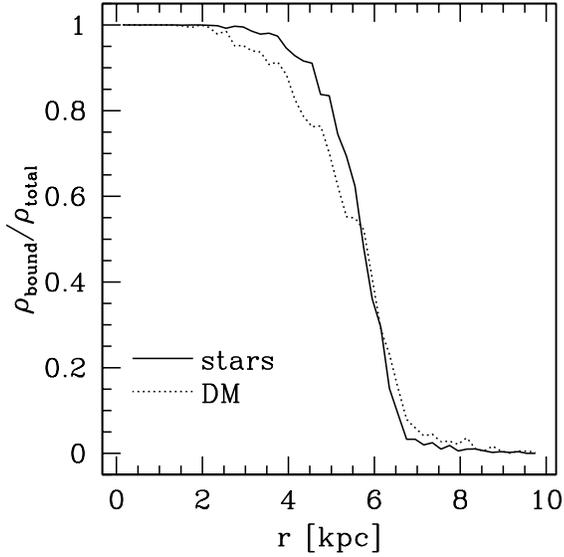}
\end{center}
\caption{Ratios of the number of bound particles to the total number of particles for stars (solid line)
and DM (dotted line) as a function of the distance from the dwarf centre.}
\label{boundratio}
\end{figure}

A similar comparison for the stellar particles is shown in the middle panel of Fig.~\ref{density}.
Again the symbols indicate respectively bound stars (open circles) and all stars (open triangles).
The dashed line is the {\em deprojected\/} S\'ersic profile (S\'ersic 1968; Prugniel \& Simien 1997)
\begin{equation}    \label{depsersic}
	\nu(r) = \nu_0 \left( \frac{r}{R_{\rm S}} \right)^{-p}
	\exp \left[-\left( \frac{r}{R_{\rm S}}
	\right)^{1/m} \right]
\end{equation}
describing the 3D density distribution of the stars, where the constants are related to the
usual surface density distribution
\begin{equation}    \label{sersic}
	\Sigma(R) = \Sigma_0 \exp [-(R/R_{\rm S})^{1/m}]
\end{equation}
by $\nu_0 = \Sigma_0 \Gamma(2 m)/\{2 R_{\rm S} \Gamma[(3-p) m]\}$ and
$p = 1.0 - 0.6097/m + 0.05463/m^2$ (see Lima Neto, Gerbal \&
M\'arquez 1999; {\L}okas 2002). Our best-fitting parameters for the density profile of
bound stars are listed in the first row of Table~\ref{proj}. The solid line in the middle
panel of Fig.~\ref{density} shows the spherically averaged exponential profile of the initial
stellar disk.

The density distributions shown in the two upper panels of Fig.~\ref{density} indicate that the dwarf
lost a significant part of both its mass components. Counting the bound particles in the final stage,
we find the final dark component has a mass of $3.3 \times 10^7$ M$_{\sun}$ and the stellar one
$1.3 \times 10^7$ M$_{\sun}$. \label{dwarfmass}
These should be compared to the initial masses which were respectively
$4.1 \times 10^9$ M$_{\sun}$ and $1.5 \times 10^8$ M$_{\sun}$. We therefore find that during the
evolution the dwarf lost 99 percent of its DM and 91 percent of the stellar mass.

\begin{figure}
\begin{center}
    \leavevmode
    \epsfxsize=7.5cm
    \epsfbox[0 0 370 370]{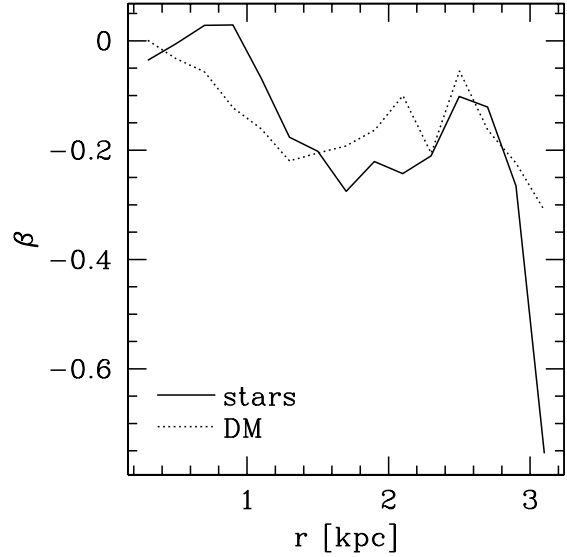}
\end{center}
\caption{Profiles of the anisotropy parameter $\beta$ for bound stellar (solid line) and dark
particles (dotted line).}
\label{plotbeta}
\end{figure}

In the lower panel of Fig.~\ref{density} we plot the mass-to-light density
ratios measured for the bound and
for all particles (again with open circles and triangles respectively). The values were expressed
in solar units assuming a mass-to-light ratio for the stars of $\Upsilon = 3$ M$_{\sun}/$L$_{\sun}$
typical for stellar populations of dSph galaxies (Schulz et al. 2002) so that the distribution
of light is $l(r) = \nu (r)/\Upsilon$ with $\nu (r)$ given by equation (\ref{depsersic}).
We note that within the main body of the dwarf ($r<2.5$ kpc) the mass-to-light density
ratio is almost constant with a mean value of 10.3 M$_{\sun}/$L$_{\sun}$. Again,
for comparison we also show with a solid line the initial ratio of mass and light densities.

It is also interesting to look at the ratios of the number of bound particles to the total number
of particles for the two components as a function of radius which are plotted in Fig.~\ref{boundratio}.
Note, that although tidal tails start to dominate the density at about 2.5 kpc from the centre of
the galaxy (see Fig.~\ref{dwarfplot}), up to about 5 kpc most of the tidal tail particles are
still bound. The fact that some of the material in the tails is bound to the satellite was
first noticed by Johnston (1998). The magnitude of this effect will however depend on the
mass of the dwarf and its position on the orbit. The situation poses a potential difficulty in
dynamical modelling: even though some stars were stripped from the
dwarf and formed the tidal tails, they still have velocities not very different from the dwarf's mean
and cannot be distinguished by an observer from stars belonging to the spheroidal component.
Although they are not good tracers
of the dwarf's potential they mimic the member stars just because they follow the dwarf on its
orbit around the MW. We address this issue further in the discussion section.

Another quantity of interest in modelling dSph galaxies is the anisotropy parameter
relating the tangential $\sigma_t$ and radial $\sigma_r$ velocity dispersions
\begin{equation}    \label{beta}
	\beta=1-\frac{\sigma_t^2(r)}{2\sigma_r^2(r)}
\end{equation}
where $\sigma_t^2 = \sigma_\theta^2 + \sigma_\phi^2$. The parameter describes the type of
orbits of a tracer in a gravitational potential with radial orbits
corresponding to $\beta=1$, isotropic to $\beta=0$ and circular orbits
to $\beta \rightarrow - \infty$. Although only the stellar anisotropy parameter can be inferred from
observations, here we plot its behaviour for both components to see if there is any significant
difference between them. Fig.~\ref{plotbeta} shows the anisotropy profiles measured for bound
stellar and dark particles with solid and dotted lines respectively. The data were averaged in radial
bins of width 0.2 kpc. Velocity dispersions were calculated directly using the standard unbiased estimator
({\L}okas et al. 2005). We conclude that the anisotropy of both stellar and DM particles is weakly
tangential as expected for tidally stripped objects (see e.g. Read et al. 2006a), but
does not depart strongly from zero. The stellar value averaged over all bins up to $r=2.5$ kpc
is $\beta=-0.13$. We also computed the anisotropy using equation (\ref{beta}) where instead of
velocity dispersions we used mean squared velocities. There are no differences with the ones of
Fig.~\ref{plotbeta} within $r<2$ kpc, indicating negligible streaming motions inside $2$ kpc.

\section{Removal of interlopers}

\subsection{Contamination from the tidal tails}

The presence of the unbound stars from the tidal tails must affect any inferences concerning the
mass content in dSph galaxies based on equilibrium models. In order to show how these stars can
influence the measured velocity dispersion profile we choose three different directions of
observation: along the tidal tails, perpendicular to it and an intermediate one (at $45^{\circ}$)
to both of the previous ones. We project all the stellar velocities along the line of sight and
the positions of the stars on the plane of the sky. The velocities are measured with respect to
the dwarf's mean velocity in a given direction and we introduce a cut-off in velocity corresponding
to approximately $4 \sigma_{\rm los}$, where $\sigma_{\rm los}$ is the line-of-sight velocity
dispersion. In this way mock catalogues of kinematic data listing positions on the sky and line-of-sight
velocities of the stars can be created.

\begin{figure}
\begin{center}
    \leavevmode
    \epsfxsize=7.5cm
    \epsfbox[0 0 370 520]{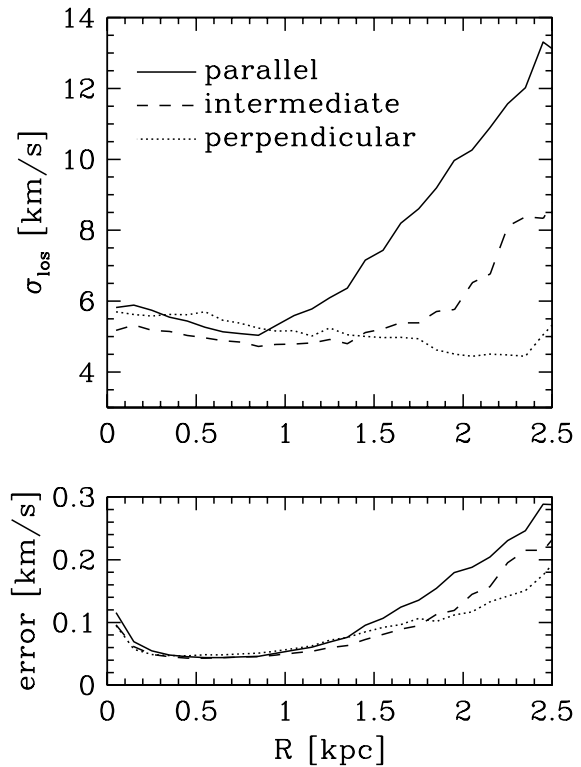}
\end{center}
\caption{Velocity dispersion profiles measured for the line of sight positioned along the tidal tails
(solid line), perpendicular to the tails (dotted line) and at an intermediate angle (dashed line).}
\label{disp}
\end{figure}

Fig.~\ref{disp} shows the velocity dispersion profiles and their errors
measured for all stars within the projected
radius $R = 2.5$ kpc from the dwarf's centre and velocities smaller than 30 km s$^{-1}$ with respect to
the dwarf's mean velocity. We again used the standard unbiased estimator of dispersion and assigned
the sampling errors of size $\sigma_{\rm los}/\sqrt{2(n-1)}$ where $n$ is the number of particles per
bin. The dispersions were calculated in bins of equal size in $R$ (i.e. with different number
of particles per bin).
The solid line plots the dispersion measured when the line of sight is
along the tidal tails, the dotted line for the line of sight perpendicular to the tails and
the dashed line for the intermediate angle. We immediately notice that, in the case when the observation
is done along the tidal tails, the velocity dispersion at $R > 1$ kpc shows a secondary increase
caused by the presence of unbound particles in the tails. A similar increase was also noticed
by Johnston, Sigurdsson \& Hernquist (1999), Read et al. (2006b) and Sohn et al. (2006) in their
simulated kinematic samples. The dispersion is not significantly altered
in the centre of the dwarf because the bound stars are numerous enough there to outnumber the unbound
ones. The effect of unbound stars is much weaker in the case of the intermediate direction of
observation and absent when the line of sight is perpendicular to the tails.

The methods to deal with such unbound objects or {\em interlopers\/} were previously
considered mainly in the context of
modelling the velocity dispersion profiles of galaxies in clusters. In that case the problem has
been recognized and studied for a long time because of larger samples available from which a
velocity dispersion profile, not only central velocity dispersion, could be measured. In addition,
it was quite obvious that many background and foreground galaxies unbound to a given cluster
are detected in a redshift survey. In the case of dwarf spheroidal galaxies, the stellar samples
have only recently become large enough to study the profiles of the velocity moments. As for the
sources of contamination, it seemed quite likely that stars from the MW can contribute, it
remained unclear however to what extent the samples can be biased by the stars stripped from the
dwarf or even whether such a process indeed occurs.

Many prescriptions were proposed in the literature on how to deal with the interloping galaxies
in clusters. The traditional approach, still widely used today, goes back to Yahil \& Vidal (1977)
who proposed to calculate the velocity dispersion $\sigma_{\rm los}$ of the sample
and then iteratively reject outliers with velocities
larger than $3 \sigma_{\rm los}$ with respect to the mean. The only way, however, to
test the efficiency and reliability of different methods is to use a set
of $N$-body simulations where the whole information about the properties of the tracer is available.
Recently Wojtak et al. (2006) performed such a study using data samples generated from 10
simulated DM haloes with masses of the order of a galaxy cluster observed in three
orthogonal directions. Among the {\em direct\/} methods of interloper removal they considered
(when the interlopers
are identified and removed from the sample before the proper dynamical analysis of the object)
the most effective turned out to be the one proposed by den Hartog \& Katgert (1996). In addition,
the method does not involve any parameters specific to clusters so is quite general, i.e. seems
to be applicable to any gravitationally bound objects (for the application to galaxy clusters see
e.g. Wojtak \& {\L}okas 2006). Here we will
therefore use only this method and test its performance in the case of dSph galaxies.

The method assumes that a tracer particle (in our case a star in the dwarf galaxy) can be either on
a circular orbit with velocity $v_{\rm cir} = \sqrt{GM(r)/r}$ or falling freely in the galaxy's
potential with velocity $v_{\rm inf} = \sqrt{2}v_{\rm cir}$ (see den Hartog \& Katgert 1996;
Wojtak et al. 2006). We look for the maximum
velocity among the two for a given projected radius $R$ and iteratively reject all stars with
larger velocities. The mass profile needed for the calculation of the velocities is estimated
in each iteration using the standard mass estimator $M_{VT}$ derived from the virial
theorem (Heisler, Tremaine \& Bahcall 1985)
\begin{equation}	\label{mass}
	M_{VT}(r=R_{\rm max})=\frac{3\pi N}{2G}\frac{\Sigma_{i}
	(V_{i}-\bar{V})^{2}}{\Sigma_{i<j}1/R_{i,j}} ,
\end{equation}
where $N$ is a number of stars with projected radii $R<R_{\rm max}$, $V_{i}$ is the line-of-sight
velocity of the $i$-th star and
$R_{i,j}$ is a projected distance between $i$-th and $j$-th star.
The mass profile can be simply obtained as $M(r)\approx
M_{VT}(R_{i}<r<R_{i+1})$, where $R_{i}$ is the sequence of projected radii of
stars in the increasing order.

\begin{figure}
\begin{center}
    \leavevmode
    \epsfxsize=7.0cm
    \epsfbox{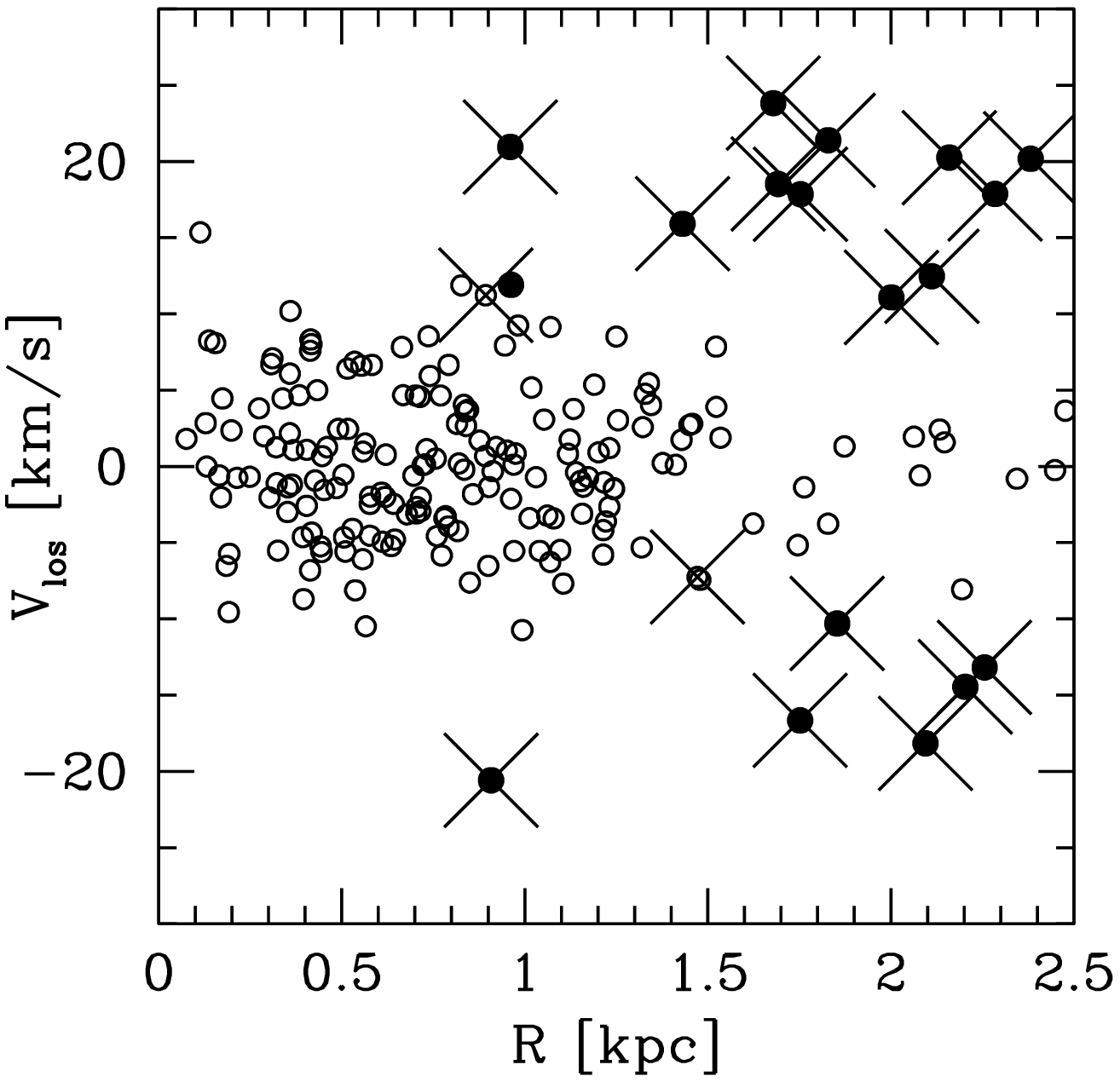}    \epsfxsize=7.0cm
    \epsfbox{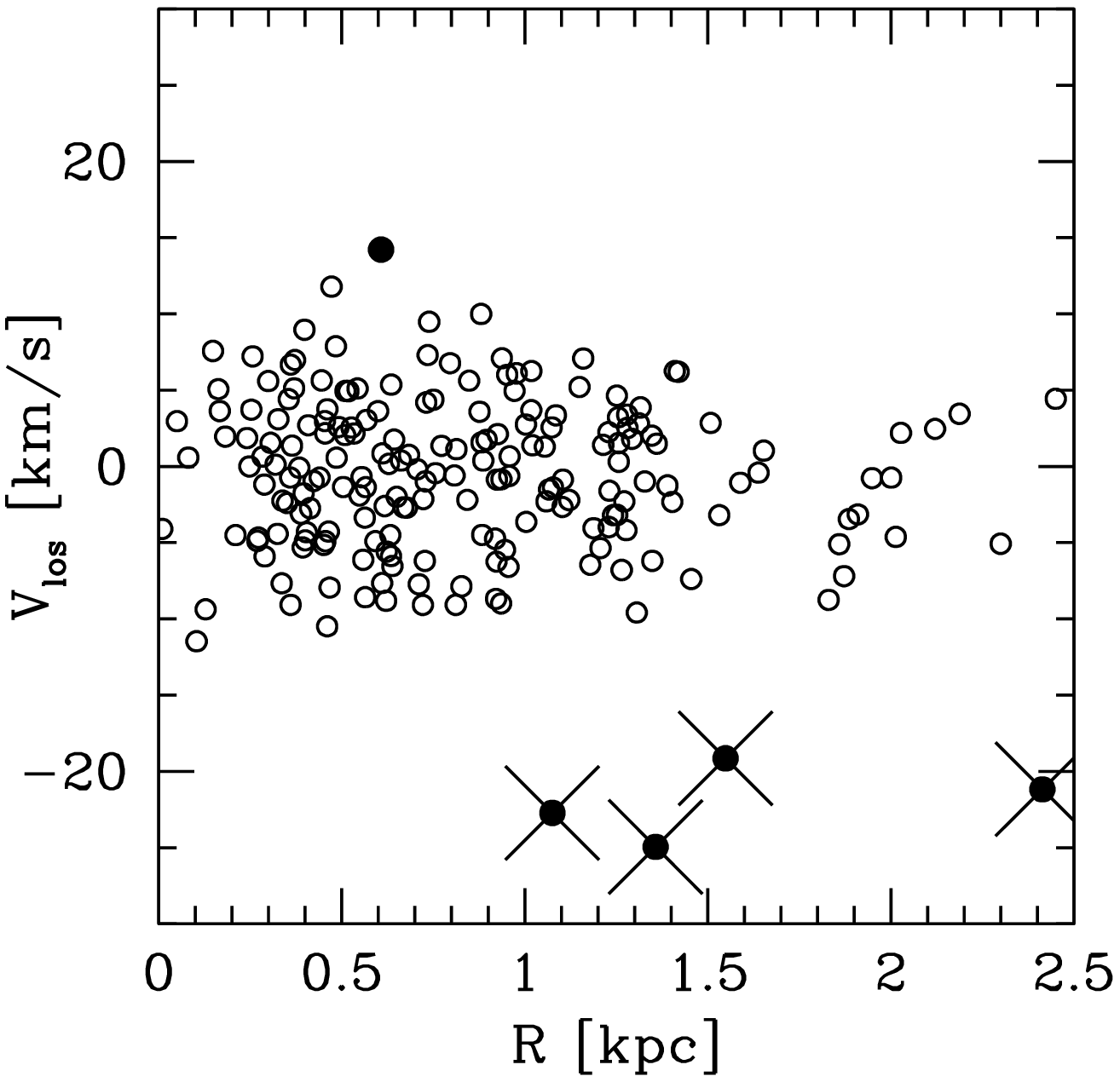}    \epsfxsize=7.0cm
    \epsfbox{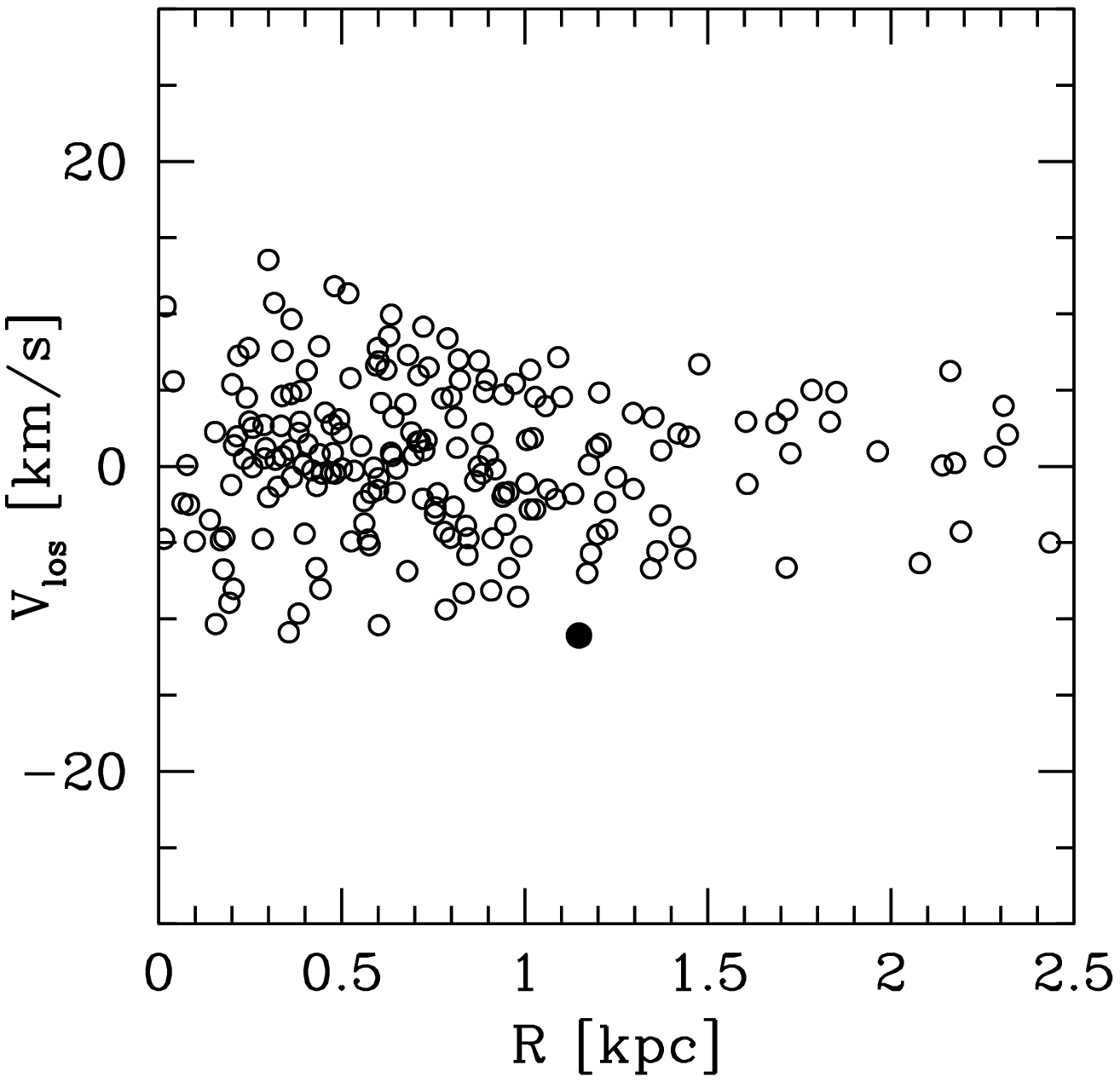}
\end{center}
\caption{Examples of velocity diagrams showing samples of 200 stars selected randomly for the
observations performed in different directions: along the tidal tails (upper panel),
perpendicular to the tails (lower panel) and at the intermediate angle (middle panel).
Unbound stars
are marked with crosses. Stars rejected as interlopers are plotted with filled circles, those accepted
by the procedure are shown as open circles.}
\label{diagrams}
\end{figure}

We construct realistic samples of kinematic data for the dwarf by randomly selecting from
our observational cylinder (with $R_{\rm max} = 2.5$ kpc and $V_{\rm max} = 30$ km s$^{-1}$)
200 stars, a number similar to the number of stars with measured velocities in
best-studied dSph galaxies at present. The data are most conveniently displayed in the form of
a {\em velocity diagram\/}, i.e. a plot of line-of-sight velocity versus projected distance from the
dwarf's centre for
each star. Three typical examples of such diagrams are shown in Fig.~\ref{diagrams}
for the three directions of observation: along the tidal tails (upper panel), at the
intermediate $45^\circ$ angle to the tails (middle panel) and in the direction perpendicular
to the tails (lower panel). In each panel crosses indicate stars identified as
unbound to the dwarf using the full 3D information. The contamination due to these stars averaged
over 100 independent samples is 9.8, 2.6 and 0.04 percent respectively for the three directions
of observation.

\begin{table}
\begin{center}
\caption{Fractions of removed interlopers ($f_i$) and member (bound) stars ($f_m$) in percent of
the total number of observed stars ($N$) averaged over 100 velocity diagrams. Numbers are given
separately for three directions of observation: along the tidal tails (a), perpendicular to it (c)
and an intermediate direction (b).}
\begin{tabular}{ccccccc}
$N$      & $f_i$(a) & $f_m$(a) & $f_i$(b) & $f_m$(b) & $f_i$(c)   & $f_m$(c)  \\
\hline
200      & 78.8  &  0.6  & 70.4  & 1.1   & 75.0    & 0.8    \\
400      & 79.4  &  0.7  & 71.4  & 1.2   & 75.8    & 1.1    \\
800      & 80.4  &  0.7  & 71.3  & 1.2   & 77.5    & 1.1    \\
\hline
\label{fractions}
\end{tabular}
\end{center}
\end{table}

In each of the samples we iteratively rejected
the interlopers using the prescription described above. The stars rejected by the procedure are marked
with filled circles while those accepted are plotted as open circles. We can see that most of the
unbound stars are correctly identified as interlopers by the procedure. There are cases however, when
bound stars are misidentified as interlopers and there are unbound stars which are not rejected.
Averaging over 100 independent velocity diagrams we
find that 70-80 percent of unbound stars are removed
from the samples, while only about 1 percent of bound stars are rejected. A similar experiment was
repeated for samples with 400 and 800 stars. The results are summarized in Table~\ref{fractions}
which proves that the procedure works similarly for samples of different size.

We also checked the dependence of the performance of
this method on the initial cut-off in velocity. Namely, we rejected
the stars with velocities differing from the dwarf's mean velocity by more than $3\sigma$
(which corresponds to about 18 km s$^{-1}$ in our case) instead of the original $4\sigma$,
and then applied our interloper removal
procedure. The results, in terms of the final cleaned samples, were in almost all (out of a hundred) cases
identical as the ones we obtained with a $4\sigma$ cut-off. This proves that the method does not
depend significantly on the initial cut-off in velocity.

We also compared the effectiveness of this method to the simple, non-iterative, cut-off in velocity
as commonly
applied to kinematic samples of dSph galaxies (e.g. Wilkinson et al. 2004; Walker et al. 2006).
In the case of a cut-off at $3\sigma$ level we reject on average only from 20 up to 55 percent
of interlopers,
depending on the angle of view. This suggests that one could try to use a more stringent rejection
criterion, e.g. with cut-off at $2.5\sigma$ (as advertised by Walker et al. in the case of Fornax).
In this case the effectiveness indeed increases to the range between
36 and 62 percent depending on the direction of observation.
Unfortunately such a stronger criterion causes significant fraction of bound stars to be rejected
while our procedure does not suffer from this problem.

\begin{figure}
\begin{center}
    \leavevmode
    \epsfxsize=7.5cm
    \epsfbox[0 0 370 370]{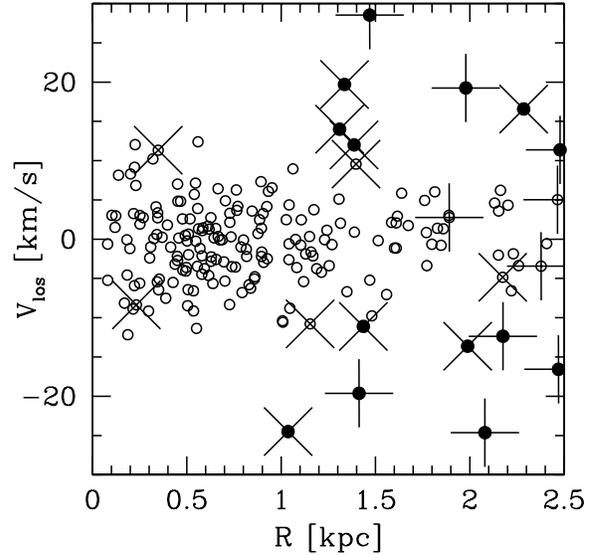}
\end{center}
\caption{An example of velocity diagram showing a sample of 200 stars including the contamination
from the tidal tails and from the MW. Unbound stars originating from the dwarf
are marked with crosses ($\times$), stars from the MW are marked with pluses ($+$).
As in Fig.~\ref{diagrams}, stars rejected as interlopers are plotted with filled circles,
those accepted by the procedure are shown as open circles.}
\label{stars}
\end{figure}

One may wonder if the good performance of our method of interloper removal is not restricted to
systems where mass follows light. We rely on the use of the virial mass estimator which has been
shown to often lead to underestimation of mass when applied to tracer populations in systems with
increasing $M/L$ ratios (Evans et al. 2003). The full answer to this question is beyond
the scope of this paper since it would require another simulation of a different system in order to
create a realistic unbound population of stars. However, in order to check whether the underestimation
of the virial mass may lead to removal of {\em bound\/} stars from the kinematic sample we generated
an equilibrium two-component system with an extended dark matter halo and mass-to-light density
ratio increasing from 7 near the centre to 20 $M_{\sun}/L_{\sun}$ at $r=3$ kpc. The dark matter profile
was given by the NFW formula with $M_{\rm vir}=10^9 M_{\sun}$ and $c=15$, while stars were distributed
according to the S\'ersic profile
with $M=10^8 M_{\sun}$, $R_{\rm S}=1$ kpc and $m=1$. The orbits were isotropic. We repeated the
procedure of generating velocity diagrams with 200 stars and applied our interloper removal scheme. We
find that on average only less than 1 percent of the stars are rejected. This encourages us to believe
that the procedure should also be applicable to systems with increasing $M/L$ ratios like Draco
(see S\'anchez-Conde et al. 2007), although its efficiency remains to be tested.

\subsection{Contamination from the MW stars}

Depending on the systemic velocity of the dwarf and its position with respect to the MW
we can also expect some contamination from the MW stars. In the kinematic samples for dSph
galaxies presently available these stars are not identified (except for Carina, see Mu\~noz et
al. 2006) and have to be treated in the same way as the interlopers originating from the dwarf
itself. To emulate such contamination we add to our simulated data a distribution of stars
originating from the MW disk and bulge.
This distribution follows the dynamical mass model A1 for the MW of Klypin et al. (2002)
described in section 2 (we have verified using the Besancon Galaxy model, described in more detail
in section 6, that the contribution of the MW stellar halo is negligible in observations of
dSph galaxies). Our
distribution of MW stars is much more coarse-grained than the distribution of stars in the
dwarf itself which mimics to some extent the observational approach of constructing kinematic
samples for dSph galaxies by selecting only stars along the red giant branch of the dwarf.

Working with
the final output of the simulation (when the dwarf is at some particular position with respect to
the MW disk) we do not have much choice in selecting the position of the
observer. The observer
is placed at 8 kpc from the MW centre along the line connecting the dwarf and the MW centre. In
this particular configuration the observer happens to look at the dwarf almost along the tidal tails.
After these modifications we repeat the mock observations as before.
An example of the velocity diagram for 200 stars obtained in such observation is shown in
Fig.~\ref{stars}. As before, the unbound stars originating from the dwarf are marked with crosses
($\times$), while stars from the MW are marked with pluses ($+$). The total contamination
of the sample is now 14.2 percent.
As in Fig.~\ref{diagrams}, stars rejected as interlopers are plotted with filled circles,
those accepted by the procedure are shown as open circles. The fraction of removed interlopers
averaged over 100 independent samples is now 66 percent (63 percent for the interlopers
from the tidal tails of the dwarf, 68 percent for the MW stars). These fractions are significantly
lower than for samples without the contamination from the MW (see Table~\ref{fractions}). The
reason for this lies probably in the completely different velocity distribution of the MW stars.

\begin{figure}
\begin{center}
    \leavevmode
    \epsfxsize=7.5cm
    \epsfbox[0 0 370 370]{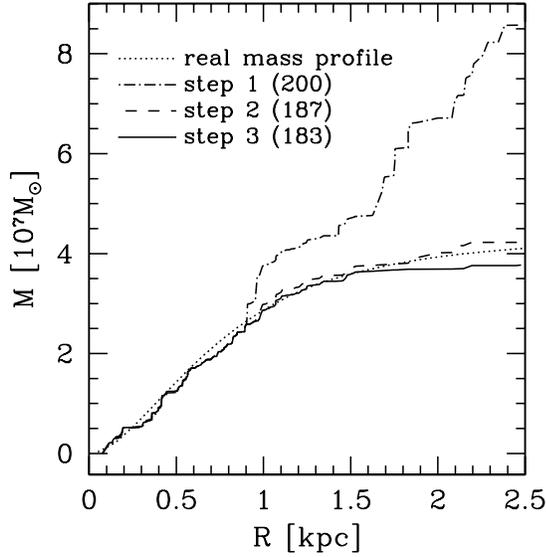}
\end{center}
\caption{Mass profiles obtained for the most contaminated sample observed along the tidal tails
from the upper panel of
Fig.~\ref{diagrams} in subsequent iterations of the procedure for interloper removal using the
mass estimator (\ref{mass}).
The dashed-dotted, dashed and solid lines show the profiles in the first, second and third iteration
respectively. The numbers of stars in each iteration are given in the legend. For comparison
we also plot with a dotted line the real mass profile of the dwarf.}
\label{massprofile}
\end{figure}

\section{Modelling of the velocity dispersion}

As is clearly seen in the upper panel of Fig.~\ref{diagrams}, the largest number of
unbound stars is present in
the samples when the line of sight is along the tidal tails. These stars are responsible for the
increasing velocity dispersion profile shown with the solid line in Fig.~\ref{disp}. The bias caused
by these stars in the mass estimate can be simply demonstrated by the evolution of the
mass profile calculated from the samples using formula (\ref{mass}). In Fig.~\ref{massprofile}
we show the mass profiles obtained in the subsequent iterations of our procedure for interloper
removal performed for the sample shown in the upper panel of Fig.~\ref{diagrams}.
We see that in the case when the contaminated sample with all 200 stars is used the total mass
is overestimated by a factor of a few. After a few iterations the mass estimate from (\ref{mass})
and the real mass profile measured from the 3D data (shown with a dotted line) converge.

\begin{figure}
\begin{center}
    \leavevmode
    \epsfxsize=7.5cm
    \epsfbox[0 0 370 370]{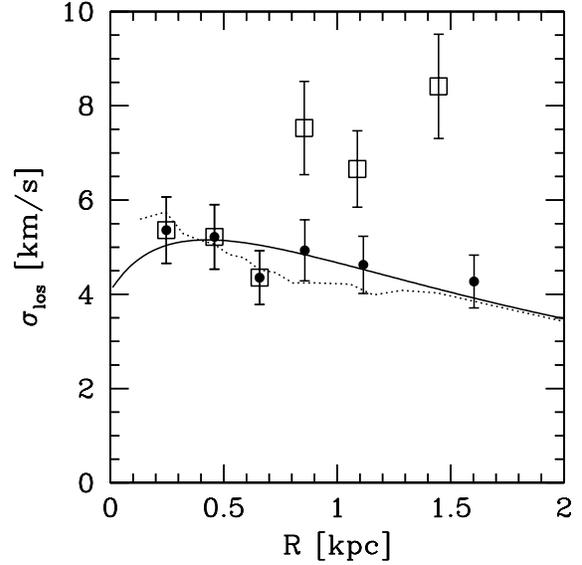}
\end{center}
\caption{Velocity dispersion profiles for the realistic sample of 200 stars observed
along the tidal tails and shown in the upper panel of Fig.~\ref{diagrams}. The points marked with
open squares with
$1\sigma$ error bars were calculated with 30 stars per bin from the total sample (without removal of
interlopers), filled circles for the cleaned sample (after removal of interlopers). The solid line
shows the best fitting solution of the Jeans equation for the cleaned sample. The dotted line represents
the velocity dispersion profile measured for all bound particles observed at this angle.}
\label{disper}
\end{figure}

Fig.~\ref{disper} shows the velocity dispersion profiles $\sigma_{\rm los}(R)$
for the sample of 200 stars from the upper
panel of Fig.~\ref{diagrams} (observation along the tidal tails).
The data points were obtained by binning the data with 30 stars per bin. We used standard unbiased
estimator to calculate the dispersions and the data points were assigned standard sampling error bars
of size $\sigma_{\rm los}/\sqrt{2(n-1)}$ where $n=30$ is the number of stars per bin.
The open squares are the velocity dispersions calculated for the total sample of 200 stars before
the interloper removal procedure was applied and the filled circles show the results after removal
of interlopers. The first three data
points overlap since no stars were removed from the inner part of the diagram (low $R$). As expected,
the contaminated sample produces a velocity dispersion profile increasing with projected radius,
while the cleaned sample shows a decreasing profile. For comparison we also show in the Figure
the velocity dispersion profile measured for all bound stellar particles in this direction (dotted
line). It is clear that our data points for the cleaned sample agree well with this profile thus
confirming that the small data set of 200 stars is adequate to reliably estimate the true dispersion.

Our simulated dSph galaxy has a mass-to-light and velocity anisotropy almost constant with
radius so the measured velocity dispersion profile can be reproduced by just two constant parameters
$M/L$ and $\beta$. Following {\L}okas et al. (2005, 2006) we will therefore fit these two parameters
by adjusting to the data a solution of the Jeans equation
\begin{eqnarray}	\label{sigma}
	\nonumber \sigma_{\rm los}^2 (R) & = & \frac{2}{\Sigma(R)} \int_{R}^{\infty}
	\frac{\nu \sigma_r^2 r}{\sqrt{r^2 - R^2}} \left( 1-\beta \frac{R^2}{r^2} \right)
	{\rm d} r\\
	 & = & \frac{2}{\Sigma(R)} \int_{R}^{\infty} \nu M K \left( \frac{r}{R} \right)
	\frac{{\rm d} r}{R}
\end{eqnarray}
(where the dimensionless kernel $K$ is given in Mamon \& {\L}okas 2005, eq.~A16, and another
form is given in Mamon \& {\L}okas 2006).
Since we assume that mass follows light the shape of the mass profile is strongly constrained and
can influence the $\sigma_{\rm los} (R)$ profile only shifting it to higher or lower values
according to the value of $M/L$. The anisotropy parameter, on the other hand, can change the shape
of $\sigma_{\rm los} (R)$ making it more increasing (for very negative $\beta$) or decreasing (for $\beta$
close to zero or positive).

\begin{figure}
\begin{center}
    \leavevmode
    \epsfxsize=7.5cm
    \epsfbox[0 0 370 370]{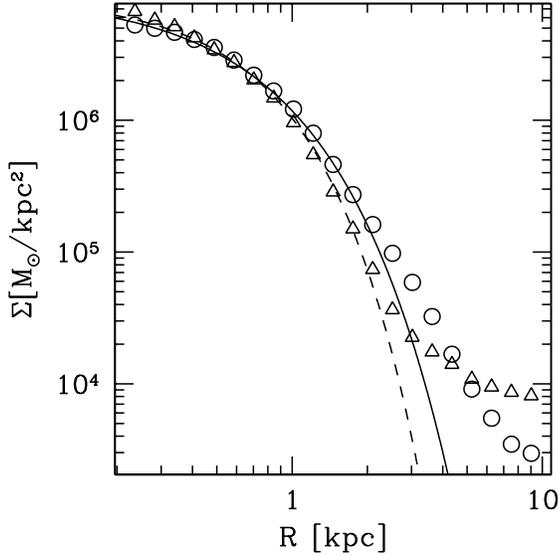}
\end{center}
\caption{The surface density distribution of stars, together with fits of formula (\ref{sersic}),
observed in two extreme directions: along the tidal tails
(circles and solid line) and perpendicular to the tails (triangles and dashed line).}
\label{surface}
\end{figure}

\begin{table}
\begin{center}
\caption{Parameters of the distribution of stars in the dwarf. The first row lists values obtained
by fitting formula (\ref{depsersic}) to the 3D data. The next three rows list values obtained from
fitting (\ref{sersic}) to the projected distribution observed in three directions:
along the tidal tails (a), perpendicular to it (c) and at an intermediate angle (b). Direction (d)
is almost along the tidal tails and was used in the observation including the contamination by
MW stars.}
\begin{tabular}{ccccc}
        & $\Sigma_0$[$10^7$ M$_{\sun}$/kpc$^2]$ & $R_{\rm S}$ [kpc]   & $m$ & $L_{\rm tot}
[10^6$ L$_{\sun}]$ \\
\hline
3D      & 1.54  &  0.41  & 0.99  &  5.3 \\
a       & 0.90  &  0.49  & 1.01  &  4.6 \\
b       & 0.99  &  0.47  & 0.89  &  4.0 \\
c       & 0.85  &  0.53  & 0.85  &  3.9 \\
d       & 1.15  &  0.40  & 1.11  &  4.6 \\
\hline
\label{proj}
\end{tabular}
\end{center}
\end{table}

In order to further emulate the observations, in fitting $\sigma_{\rm los} (R)$ profiles we do not
use the stellar profile found from the full 3D information in section~3 but estimate it independently
for each direction of observation from the surface distribution of stars. First, we count the stars
in the observation cylinder with $R<2.5$ kpc and translate their mass to light using the stellar
mass-to-light ratio $\Upsilon=3$M$_{\sun}$/L$_{\sun}$ to estimate the total luminosity of the galaxy
$L_{\rm tot}$. We then fit the parameters of the S\'ersic profile (\ref{sersic}) $\Sigma_0$,
$R_{\rm S}$ and $m$ and verify that by integrating the profile (\ref{sersic}) we reproduce
the value of $L_{\rm tot}$ measured by counting the stars.
Only for ideally spherical objects the resulting parameters would be exactly the same in each
direction and identical to those obtained before from the 3D analysis.

\begin{figure}
\begin{center}
    \leavevmode
    \epsfxsize=7.5cm
    \epsfbox{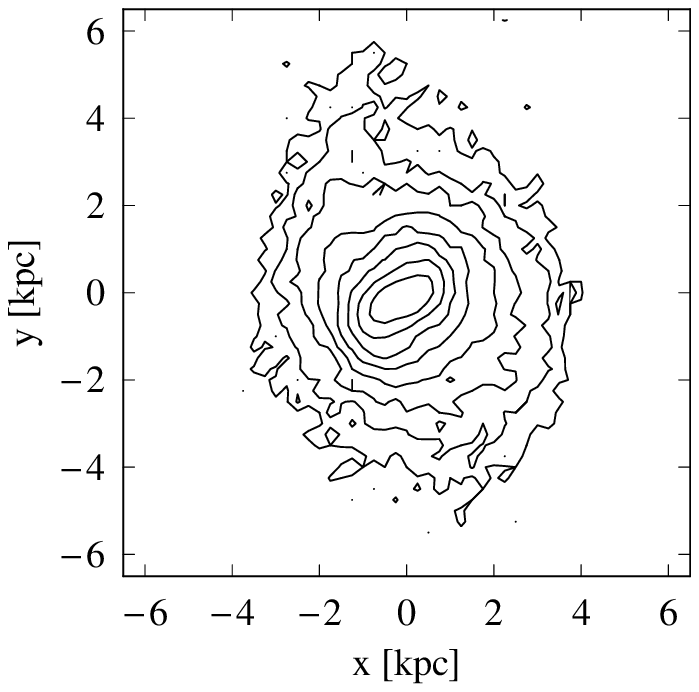}
    \epsfxsize=7.5cm
    \epsfbox{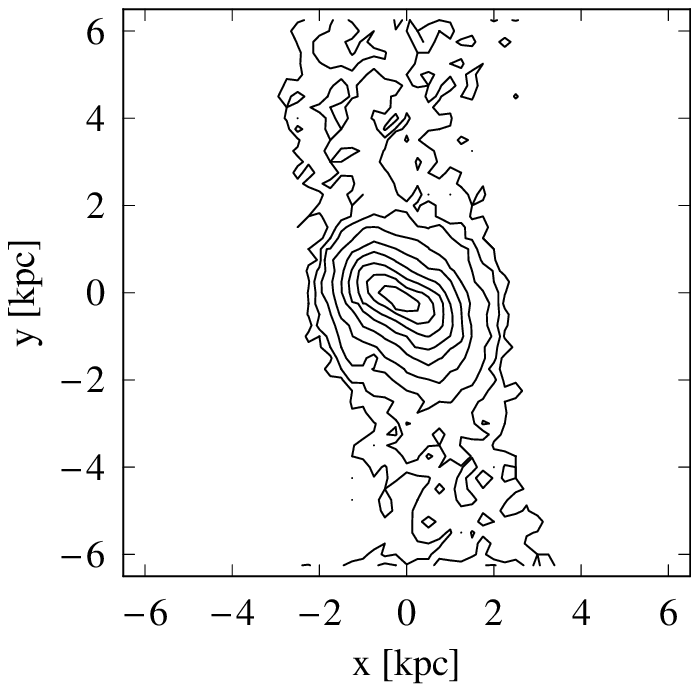}
\end{center}
\caption{Contours of equal surface density of stars in the simulated dwarf observed along the tidal tails
(upper panel) and in the direction perpendicular to the tails (lower panel). The dwarf appears
significantly more extended if observed along the tidal tails.}
\label{contours}
\end{figure}

We demonstrate in Fig.~\ref{surface} that this is indeed
not the case. The Figure compares the profiles of the projected distribution of stars
observed in two extreme directions: along the tidal tails (circles) and perpendicular to the tails
(triangles). We can see that when viewed along the tidal tails, the surface density falls more like
a power-law to larger distances and the tidal tails are not well visible. On the other hand, when
viewed perpendicular to the tidal tails the profile steepens faster but then flattens rather quickly
because the tidal tails lie in the plane of the sky. This result can be viewed as a warning against
using the shape of the stellar surface density profile to determine the tidal radius of a galaxy.
Fig.~\ref{surface} proves that the estimate of its value could be very different depending on the
orientation of the line of sight with respect to the tidal tails.

This behaviour is also illustrated by Fig.~\ref{contours}, where we can see
surface density contours for both directions of observation. Note that the scales and contour levels
are the same in both panels which means that observing the dwarf along the tails we see it as much
more extended. Fig.~\ref{surface} also plots the
fits of the S\'ersic surface density profile (\ref{sersic}) to the data restricted to $R<2$
kpc. The parameters of the fits are summarized in Table~\ref{proj}. In addition to the cases shown
in Fig.~\ref{surface} (a and c) the Table also gives the parameters for the intermediate direction of
observation (b) and the direction for which the MW stars were included (d). The values obtained
from the 3D data are listed in the first row of the Table. In the last column of the Table we
list the total luminosity of the dwarf obtained by integrating the S\'ersic density distribution
(\ref{sersic}) with these parameters and dividing by stellar mass-to-light ratio $\Upsilon$. Note that
in agreement with the impression we get from Fig.~\ref{contours} the estimated total luminosity is
largest when the observation is done along the tidal tails.

\begin{figure}
\begin{center}
    \leavevmode
    \epsfxsize=7.8cm
    \epsfbox[0 110 400 1120]{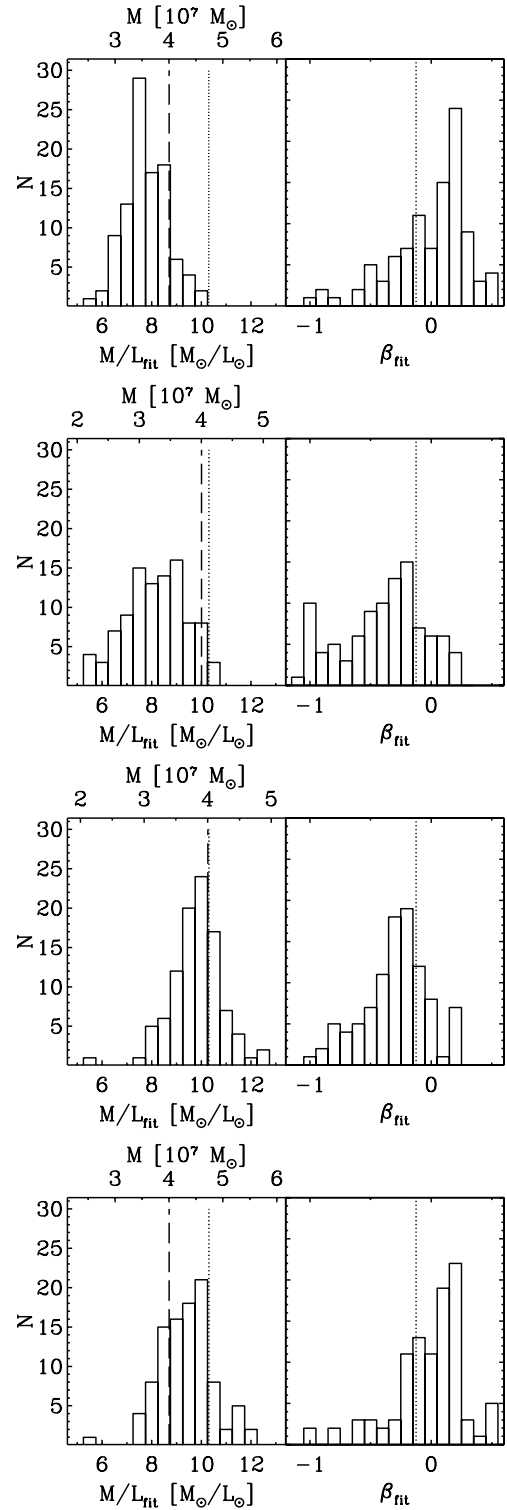}
\end{center}
\caption{Distributions of the fitted values of the parameters $M$, $M/L$ (left column) and $\beta$
(right column). The first three rows correspond respectively to three directions of
observation for the dwarf contaminated only by its own stripped stars: along the tidal tails,
at an intermediate direction and perpendicular to tidal tails. The last row shows results when the
contamination from the MW is included. The histograms on the left show results in terms of the total mass
(upper axis) and $M/L$ ratio (lower axis). Vertical lines represent true values of the parameters
of the simulated dwarf:
dotted lines of $M/L$ and $\beta$ and dashed lines of mass enclosed within the radius of 2.5 kpc.}
\label{histograms}
\end{figure}

The best-fitting solution (\ref{sigma}) to the velocity dispersion profile measured
along the tidal tails for the cleaned sample
shown in Fig.~\ref{disper} is also plotted in the Figure with a solid line.
The best-fitting parameters are: $M/L = 8.4$ M$_{\sun}/$L$_{\sun}$, $\beta = -0.25$
with $\chi^2/N=2.8/4$.
If contaminated samples were used instead, the estimated anisotropy parameter would be biased
towards lower values or a much more extended dark matter distribution would be inferred
(see {\L}okas et al. 2005). For example, for the contaminated sample in Fig.~\ref{disper}
the best-fitting parameters would be $M/L = 11.7$ M$_{\sun}/$L$_{\sun}$, $\beta = -34.8$
with $\chi^2/N=17.8/4$. This means that with the assumption of mass following light the
mass-to-light ratio is not strongly affected by contamination but strongly tangential orbits
are preferred. However, the fit quality is very bad and in order to improve it one would have
to introduce an extended dark component. This is the typical behaviour in the case of all our
contaminated samples which illustrates well the necessity to clean the samples of
interlopers before the proper modelling.

We proceed to fit the velocity dispersion profiles for our 100 randomly sampled
velocity diagrams cleaned of interlopers. The fits were done for the three directions of observation
considered before in the case of contamination by tidally stripped stars and one direction
including the contamination from the MW. The results are shown in Fig.~\ref{histograms} in terms
of the histograms of the fitted parameters $M_{\rm fit}$, $(M/L)_{\rm fit}$ and
$\beta_{\rm fit}$. Note that we really fit only two parameters: mass and anisotropy. We choose
to express the mass as a parameter by itself or in terms of the mass-to-light ratio because the latter is
affected also by the error in the estimated luminosity due to projection effects.
The left column corresponds to the fitted mass (upper axis of each histogram),
or fitted mass-to-light ratio (lower axis). The right column corresponds to anisotropy parameter.
The vertical dotted lines in each histogram show the true values of the parameters which we
adopted as the mean values of $\rho(r)/l(r)=10.3$ M$_{\sun}/$L$_{\sun}$
(shown as a dashed line in the lower panel of Fig.~\ref{density}) and $\beta = -0.13$
(Fig.~\ref{plotbeta}) averaged over distances up to $r=2.5$ kpc. The dashed line in the
left column panels indicates the total mass enclosed within the radius of 2.5 kpc from the
centre of the dwarf, $M=4 \times 10^7 M_{\sun}$. (The total mass of all bound
particles estimated in section~\ref{dwarfmass} was $M_{\rm tot}=4.6 \times 10^7 M_{\sun}$,
but we restrict ourselves here to the inner 2.5 kpc where the dwarf is almost spherically symmetric.)

\begin{table}
\begin{center}
\caption{Parameters fitted to the velocity dispersion profiles, the mass-to-light ratio $M/L$, mass $M$
and anisotropy $\beta$, averaged over 100 stellar samples cleaned of interlopers for observations
performed in three directions:
along the tidal tails (a), at an intermediate angle (b), perpendicular to tidal tails (c) and
almost along the tidal tails including the contamination by MW stars (d). The first row lists
values obtained from the 3D data.}
\begin{tabular}{cccc}
        & $M/L_{\rm fit}$ [M$_{\sun}$/L$_{\sun}$] &$M$ [$10^7 M_{\sun}$]& $\beta_{\rm fit}$ \\
\hline
3D      & $10.3$           & $4.0$ & $ -0.13 $           \\
a       & $7.56 \pm 0.88 $ & $3.5 \pm 0.4 $ & $ -0.02 \pm 0.30 $  \\
b       & $8.02 \pm 1.17 $ & $3.2 \pm 0.5 $ & $ -0.46 \pm 0.44 $  \\
c       & $9.69 \pm 1.00 $ & $3.7 \pm 0.4 $ & $ -0.33 \pm 0.26 $  \\
d       & $9.40 \pm 1.13 $ & $4.3 \pm 0.5 $ & $ 0.02 \pm 0.27 $  \\
\hline
\label{fit}
\end{tabular}
\end{center}
\end{table}

The mean values of the fitted parameters found for each case together with their dispersions
are listed in Table~\ref{fit}. Comparing to the true values measured from the 3D data (in the
first row of the Table) we find that the $M/L$ ratios are reasonably well reproduced
with slight bias towards lower values. The agreement with the true values is almost perfect
in the case when the direction of observation is perpendicular to the tidal tails.
Fitted masses (which can be obtained from $M/L$ ratios by multiplying
them by the corresponding $L_{\rm tot}$ from Table~\ref{proj}) are free from the luminosity
measurement error, but are also biased towards lower values (except for the last case).

It should be kept in mind that this bias could have several sources:
\begin{enumerate}
\item the kinematic properties of the stars observed in different directions,
\item the total luminosity measured by the observer which is systematically higher for lines of sight
more parallel to the tidal tails (see Table~\ref{proj}),
\item influence on resulting dispersion of interlopers which were not
rejected because their velocities are close to the dwarf's mean.
\end{enumerate}

\begin{figure}
\begin{center}
    \leavevmode
    \epsfxsize=7.5cm
    \epsfbox[0 0 370 520]{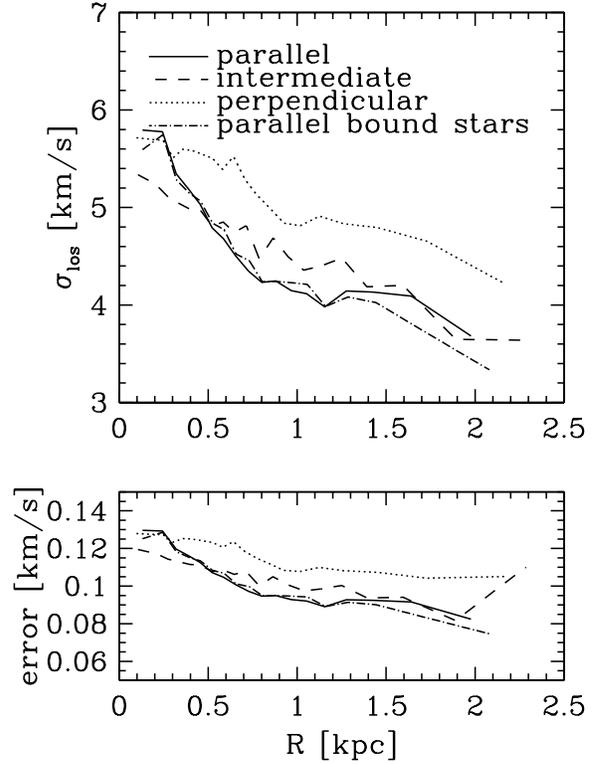}
\end{center}
\caption{Velocity dispersion profiles for combined cleaned samples observed in different directions:
along the tidal tails (solid line), perpendicular to the tails (dotted line) and at an intermediate angle
(dashed line).
Dot-dashed line corresponds to the dispersion profile for all bound stars measured in the direction parallel
to the tidal tails.}
\label{compileddisp}
\end{figure}

In order to study the source of this underestimation, for each angle of view we combined
all cleaned samples into single diagrams and measured their velocity dispersions. The results are
plotted in Fig.~\ref{compileddisp}, together with the dispersion profile for all bound particles measured
in the direction parallel to the tidal tails. It can be clearly seen that the interloper removal
scheme works perfectly well since both dispersion profiles measured along the tidal
tails overlap almost exactly. There is also a systematic shift between measurements obtained in different
directions. The dispersion profile measured in the direction parallel to the tidal tails is lower by
approximately 0.5 km s$^{-1}$ compared to the case when the dwarf is observed perpendicular to the tails,
which roughly corresponds to the difference in estimated masses. It is therefore clear
that the kinematic properties of stars are responsible for the underestimation of mass rather
than the interlopers which remained in the samples. Their influence is much smaller and according to
Fig.~\ref{compileddisp} significant only in the outer parts of the dwarf.

Another factor which can influence the results of the fitting procedure is the density profile of the
tracer which enters equation (\ref{sigma}) both as the surface distribution and the deprojected one.
In order to check its influence we repeated the fitting of the data obtained along the tails using the
surface distribution of stars seen in the direction perpendicular to the tidal tails (case (c) in
Table~\ref{proj}) which is not extended due to the tidal tails. This steeper profile (see the dashed line
in Fig.~\ref{surface}) causes the mass to be lower for most samples by about 10 percent. Therefore the
extended surface distribution along the tails that the observer would really have to use increases mass
and makes up to some extent for the lower velocity dispersion measured in this direction.
Our results in Fig.~\ref{histograms} and Table~\ref{fit} show that the combined effect of the two factors
is to underestimate the mass, i.e. the lowered velocity dispersion along the tails is more important than
the extended distribution of stars in this direction. The lower dispersion along the tails is probably due
to the fact that stars with large velocities in this direction have already escaped from the dwarf.

\section{Application to the Fornax dSph galaxy}

\begin{figure}
\begin{center}
    \leavevmode
    \epsfxsize=7.5cm
    \epsfbox[0 0 370 370]{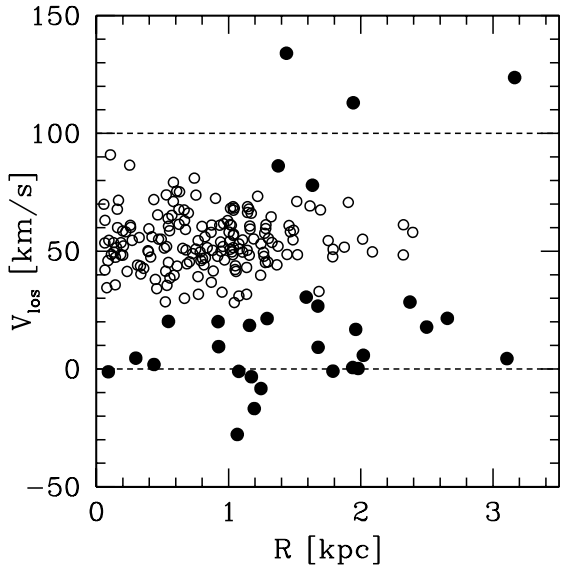}
\end{center}
\caption{Line-of-sight velocities as a function of the distance from the dwarf centre for the
sample of 202 stars from Walker et al. (2006). Dashed lines mark the initial velocity cut-off.
Filled circles mark the stars rejected by our procedure of interloper removal. Open circles
are the stars adopted as members of the dwarf.}
\label{fornaxhartog}
\end{figure}

The Fornax dSph galaxy is one of the best studied dwarfs in the vicinity of the MW. It is
believed not to be strongly dark-matter dominated with $M/L$ of the order of 10
M$_{\sun}$/L$_{\sun}$ in the range of distances constrained by kinematic data ({\L}okas
2001, 2002; Walker et al. 2006). Fornax
therefore seems to be a good representative of the class of objects modelled by our
simulation. Recently new kinematic data for Fornax stars were published by Walker et al.
(2006). Here we use this data set to apply our modelling methods of the previous sections
to a real galaxy.

The heliocentric velocities versus distance from the Fornax centre for the total sample of
202 stars from Walker et al. (after combining velocities for multiple measurements) are shown
in Fig.~\ref{fornaxhartog}. We adopted as the centre of the dwarf the coordinates
RA=$2^{\rm h} 40^{\rm m} 04^{\rm s}$, Dec=$-34^{\circ} 31^{\rm m} 00^{\rm s}$ (J2000) estimated
by Walcher et al. (2003). These coordinates are close to those estimated for the centre by
Coleman et al. (2005) but differ by a few arcmin from those by Irwin \& Hatzidimitriou (1995).
The reason for this (as discussed in detail by Coleman et al.) is that the inner and outer
iso-density contours of the distribution of stars in Fornax are not exactly concentric which
may indicate a slightly perturbed state of the object. The distances from the galaxy centre
were calculated assuming the distance of Fornax is $d=138$ kpc (Mateo 1998).

\begin{figure}
\begin{center}
    \leavevmode
    \epsfxsize=7.5cm
    \epsfbox[0 0 370 370]{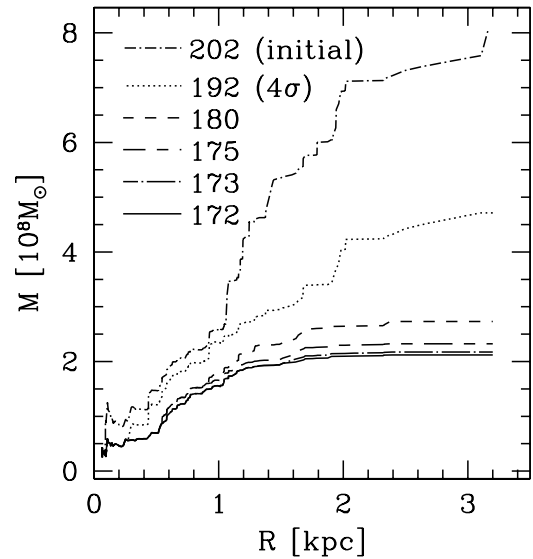}
\end{center}
\caption{Mass profiles obtained in subsequent iterations of the procedure for interloper removal
in Fornax.}
\label{mvt_hartog_cut}
\end{figure}

A preliminary selection of stars was made by choosing those with
velocities $\pm 50$ km s$^{-1}$ with respect to the sample mean of 50 km s$^{-1}$, which
corresponds to approximately $\pm 4\sigma_{\rm los}$. These stars
lie between the two horizontal dashed lines in Fig.~\ref{fornaxhartog}. The procedure for
interloper removal described in section~4 was then applied to this restricted kinematic sample.
The stars rejected as interlopers after a few iterations are marked with filled circles, those
accepted as members are plotted with open circles. The mass profiles calculated from equation
(\ref{mass}) in the subsequent iterations of the procedure are plotted in Fig.~\ref{mvt_hartog_cut}.
The final total mass from the last iteration is $2.1 \times 10^8$ M$_{\sun}$.

\begin{figure}
\begin{center}
    \leavevmode
    \epsfxsize=7.5cm
    \epsfbox[0 0 370 370]{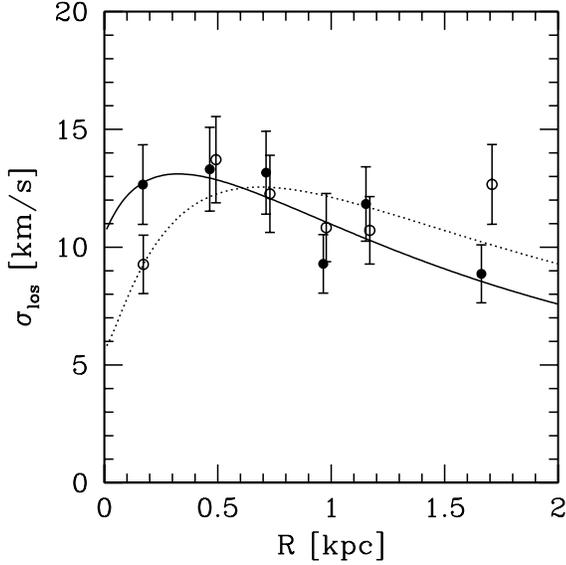}
\end{center}
\caption{The velocity dispersion profile for the cleaned
sample of the Fornax dwarf (filled circles). The solid line shows the best-fitting solution of the
Jeans equation.
Open circles and dotted line correspond to the sample obtained with 2.58$\sigma$ cut-off in velocity
(Walker et al. 2006).}
\label{fornaxdk}
\end{figure}

The velocity dispersion profile calculated from our final cleaned sample of 172 stars (using
29 stars per bin, except for 27 in the last one) is shown in Fig.~\ref{fornaxdk} as filled circles.
We fitted to this profile the solutions of the Jeans equation, as described in section~5, assuming
constant $M/L$ and $\beta$. The adopted parameters of the distribution of the stars
are listed in Table~\ref{fornax}. They are taken from Irwin \& Hatzidimitriou (1995) except for the
total luminosity which was adjusted to the different value of the distance adopted here. The Table
also gives the total mass in stars obtained assuming the stellar mass-to-light ratio $\Upsilon_V
= 3$ M$_{\sun}$/L$_{\sun}$.

\begin{figure}
\begin{center}
    \leavevmode
    \epsfxsize=7.5cm
    \epsfbox[0 0 370 370]{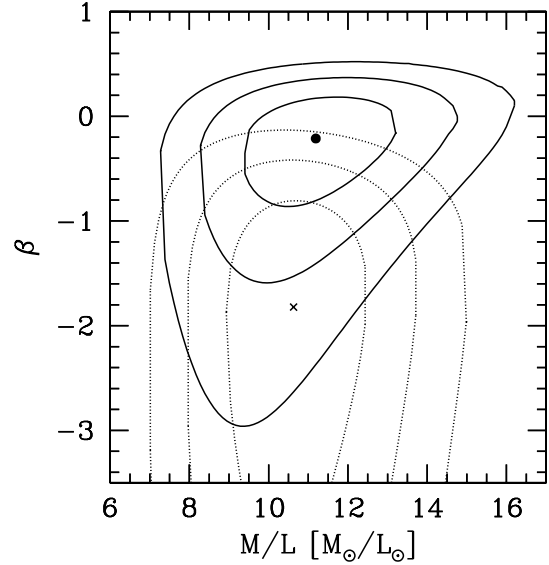}
\end{center}
\caption{The $1\sigma$, $2\sigma$ and $3\sigma$ confidence regions (solid lines) in the $M/L-\beta$ parameter
plane from fitting $\sigma_{\rm los}$ profile for Fornax. Dot indicates the best-fitting parameters.
Cross and dotted lines
correspond to the sample selected by $2.58 \sigma$ cut-off in velocity (Walker et al. 2006).}
\label{fornaxcont}
\end{figure}

The best-fitting parameters from the fitting of the velocity dispersion are
$M/L=11.3^{+2.1}_{-1.8}$ $M_{\sun}/L_{\sun}$, $\beta=-0.17^{+0.37}_{-0.63}$ with $\chi^2/N=3.4/4$.
The $1\sigma$, $2\sigma$ and $3\sigma$ confidence regions in the
parameter plane following from the sampling errors of the dispersion
are shown in Fig.~\ref{fornaxcont} with solid lines, while the best-fitting dispersion profile is shown
in Fig.~\ref{fornaxdk} as a solid line.

\begin{table}
\caption{Adopted parameters of the Fornax dwarf. }
\label{fornax}
\begin{center}
\begin{tabular}{ll}
parameter & value \\
\hline
luminosity $L_{\rm tot, V}$              & $1.8 \times 10^7 L_{\sun}$ \\
distance $d$                             & 138 kpc \\
stellar mass-to-light ratio $\Upsilon_V$ &  3 M$_{\sun}$/L$_{\sun}$  \\
stellar mass $M_{\rm S}$                 & $5.4 \times 10^7$ M$_{\sun}$ \\
S\'ersic radius $R_{\rm S}$              & 9.9 arcmin  \\
S\'ersic parameter  $m$                  & 1.0  \\
\hline
\end{tabular}
\end{center}
\end{table}

For comparison we also show in Fig.~\ref{fornaxdk} the velocity
dispersion data (open circles) in the case when the selection of 172 stars is made via a simple
$2.58 \sigma$ cut-off (Walker et al. 2006).
The best-fitting parameters for this sample are $M/L=10.6^{+1.8}_{-1.7}$ $M_{\sun}/L_{\sun}$,
$\beta=-1.82^{+1.02}_{-2.66}$
with $\chi^2/N=4.3/4$. The corresponding best-fitting velocity dispersion profile and probability
contours in the parameter plane are shown in Fig.~\ref{fornaxdk} and \ref{fornaxcont} with dotted lines.
We therefore confirm that in this case the isotropic models with mass following light are rejected
as found by Walker et al. (2006).
The difference in the estimated parameters is due to the different shape of the dispersion profile since
we have kept all our other assumptions the same. In agreement with our conclusions in section 4.1,
the simple cut-off in
velocity does not reject enough interlopers in the outer part while rejecting too many bound stars in the
centre. This situation manifests itself in decreasing the value of the innermost dispersion data point
and increasing the value of the outermost dispersion point compared to our values.

We have also verified how the values are affected if we fit {\em our\/} velocity dispersion profile using
instead of the S\'ersic profile for the tracer the best-fitting King profile as done by Walker et al.
The results are again shifted (although not so strongly) towards more tangential orbits which is
understandable since the King profile fit drops faster than the S\'ersic fit at large distances
from the Fornax centre. It should be noted however,
that according to fig. 2 of Irwin \& Hatzidimitriou (1995) the S\'ersic profile fits the outer
distribution of stars better than the King profile and it is this part of
the distribution which is important here.


Coming back to our best estimates, it is interesting to note that
the total mass of the Fornax dSph obtained by multiplying our best-fitting $M/L$
by the total luminosity of the galaxy, $M=2.1 \times 10^8$ M$_{\sun}$, is in excellent agreement with
the value estimated in the last iteration of the interloper removal procedure from (\ref{mass}).
This may suggest that, in the case of dSph galaxies not strongly dominated by DM,
the estimator (\ref{mass}) is a reliable measure of the mass. This is also confirmed by the
comparison between the mass profiles from (\ref{mass}) and the real mass profile for the
simulated dwarf in Fig.~\ref{massprofile}.

We now address the question of the origin of the contamination in Fornax. As discussed above, introducing
the initial cut-off in velocity and performing our procedure of interloper removal, we rejected 30 out
of 202 stars. It is worth noting that the distribution of the rejected stars is highly asymmetric:
while only 5 stars with velocities larger than the Fornax mean velocity of about 50 km s$^{-1}$
are removed, 25 interlopers are found with velocities below the mean. Since the tidal tails of
Fornax are visible in photometric studies (Coleman et al. 2005) we conclude that we are probably
observing the dwarf in the direction perpendicular or at least at some angle to the tidal tails.
Therefore the
contamination from them should not be large and in any case it should be symmetric with respect to
the galaxy mean velocity. This leads us to suppose that the majority of the contamination in Fornax
comes from the MW stars.

\begin{figure}
\begin{center}
   \leavevmode
    \epsfxsize=7.5cm
    \epsfbox{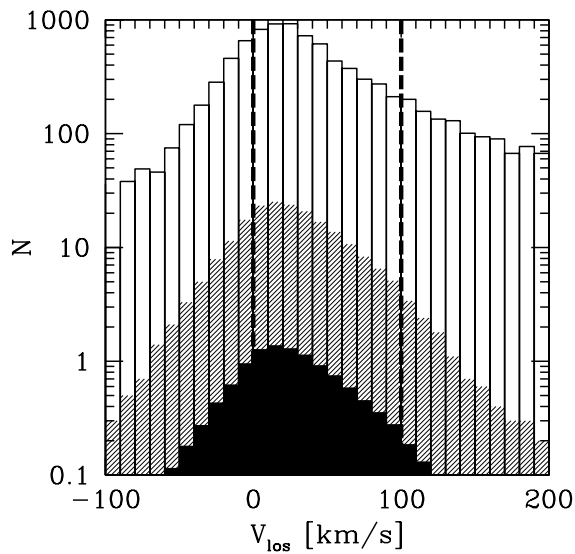}
\end{center}
\caption{Expected number of MW stars in the direction of the Fornax dSph for a field of 4 square arcminutes
in a given range of the line-of-sight velocities. The open histogram corresponds to all stars while
the half-filled one to the red giant branch in the colour-magnitude diagram in Walker et al. (2006),
from which the candidate stars for spectroscopic studies were selected. The filled histogram gives
the actual expected numbers of interlopers for the spectroscopic sample of 202 stars. Dashed lines
correspond to $4\sigma_{\rm los}$ velocity range for the Fornax sample.}
\label{besancon}
\end{figure}

Although MW contamination can be dealt with using photometric techniques (Majewski et al. 2000a), it
has not yet been done for Fornax. In order to verify our hypothesis we have therefore resorted to the
Besancon Galaxy model (Robin et al. 2003, http://bison.obs-besancon.fr/modele) which has proven
useful before to estimate the expected contamination by MW stars in the Draco dSph
({\L}okas et al. 2005).
Fig.~\ref{besancon} shows expected contamination with MW stars in the direction of the Fornax dSph
with the higher open histogram plotting the distribution of all MW stars and the middle half-filled one
the stars falling in the range of the red giant branch of the dwarf (Walker et al. 2006).
We see that the distribution of MW stars peaks at around $V_{\rm los} = 20$ km s$^{-1}$ which is where
the majority of interlopers identified in Fornax appear (see Fig.~\ref{fornaxhartog}, the mean velocity
of the 30 removed stars is 24.5 km s$^{-1}$).

>From Fig.~\ref{besancon} we expect as many as about 216 stars from the MW to lie within the red giant
branch of Fornax and about 154 of them to fall within $4\sigma_{\rm los}$ boundaries for Fornax
which correspond
to line-of-sight velocities between 0 and 100 km s$^{-1}$. Using the colour-magnitude diagram for Fornax
(fig.~1 of Walker et al. 2006) we estimate the total number of visible stars in the selected region
to be around 4000. This means that the MW contamination could be at the level of $5.5$ percent
with the distribution given by the lower filled histogram of Fig.~\ref{besancon}. With this level of
contamination we expect about 10 stars from the MW in the sample of 202 stars for Fornax.
Note that this number can only indicate an order of magnitude while the real contamination can
be different. Our calculation certainly demonstrates however that the contamination by MW stars in
the Fornax sample is significant.

\begin{figure}
\begin{center}
    \leavevmode
    \epsfxsize=7.5cm
    \epsfbox{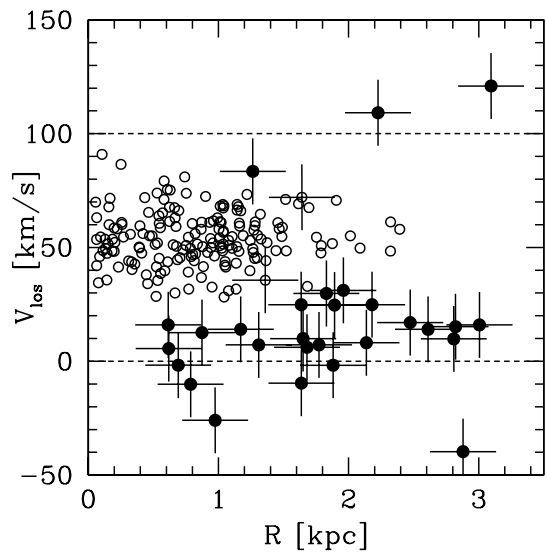}
\end{center}
\caption{An example of the velocity diagram showing a sample of 202 stars including 172 previously
selected Fornax stars as true members and 30 contaminating stars
from the MW generated with the Besancon model (marked with pluses $+$).
Stars rejected by the method of interloper removal are shown as filled circles, those accepted with
open circles.}
\label{diabes}
\end{figure}

In order to check whether this contamination can be dealt with using our method of interloper
removal, we assumed that the cleaned sample for Fornax (172 stars marked with open circles in
Fig.~\ref{fornaxhartog}) are the true galaxy members. Assuming further that all the contaminating
stars come from the MW, we generated 100 samples of 30 contaminating MW stars (so the total sample
has 202 stars, exactly as the original Fornax sample) using the Besancon model. An example of this
artificially generated contamination is shown in Fig.~\ref{diabes} where the MW stars were marked
with pluses ($+$). As expected, they cluster at around $V_{\rm los}=20$ km s$^{-1}$.
On such samples we performed the procedures of interloper removal. The results
are coded as before with open circles for the accepted stars and filled circles for the rejected ones.
The efficiency of the method in rejecting the MW stars averaged over 100 velocity diagrams is about
80 percent while only 0.5 percent of the `true' members are removed. The result is of course not
general and will vary depending on the position of the dwarf galaxy on the sky and its heliocentric
velocity. This example shows however that even in cases
where the mean velocities of the dwarf and the MW stars are not strongly separated our method is
efficient in removing the contamination.

\section{Discussion}

We have studied the origin and influence of unbound stars in kinematic samples of dSph galaxies.
Using a high resolution $N$-body simulation of a dwarf galaxy orbiting in a MW potential we
generated mock stellar samples for the galaxy observed in different directions. We find that
when the line of sight of the observer is along the tidal tails the samples can be strongly
contaminated by unbound stars leading to a secondary increase of the galaxy velocity dispersion
profile. We proposed a method to remove such interloper stars from the samples and tested
its efficiency. We compared the performance of this method to the usually applied schemes based
on the rejection of stars with velocities larger than $3\sigma$ with respect to the galaxy's mean
velocity. Such schemes turn out to be insufficient: samples thus created still show a secondary
increase in the velocity dispersion profile at larger distances from the galaxy centre
(Kleyna et al. 2002; Walker et al. 2006) leading to biased inferences concerning the mass distribution
and anisotropy. The mass distribution appears more extended and the orbits more tangential (see the
discussion in {\L}okas et al. 2005).

Recently, McConnachie, Pe\~narrubia \& Navarro (2006)
proposed that velocity dispersion profiles with such features can be interpreted in terms of additional
components in the mass distribution. Our results for the simulated dwarf as well as for Fornax
(see also S\'anchez-Conde et al. 2007 for Draco) argue against this interpretation: once the
interlopers are removed the velocity dispersion profiles of dSph galaxies are well-behaved,
decreasing functions of radius and can be modelled by smooth stellar and DM distributions.
We agree with Read et al. (2006b) that tidal stripping should not be important in the inner parts of
dSph galaxies but in the outer parts it can significantly affect the observed velocity
dispersion profile.

\begin{figure}
\begin{center}
    \leavevmode
    \epsfxsize=7.5cm
    \epsfbox{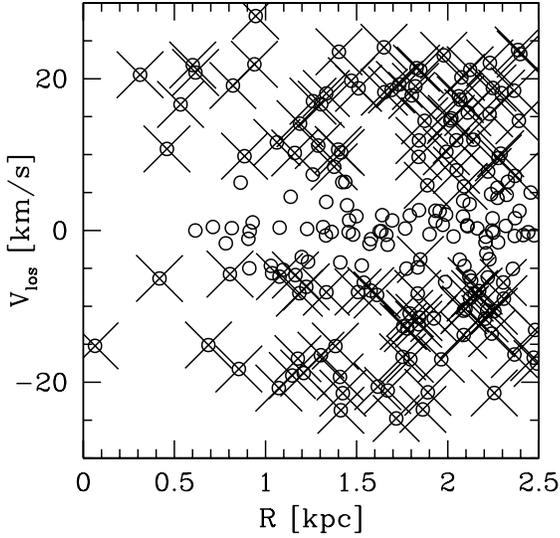}
\end{center}
\caption{An example of the velocity diagram for 200 stars originating from the tidal tails,
viewed along the tidal tails (stars with $r>2.5$ kpc). The stars unbound to the dwarf were additionally
marked with crosses.}
\label{tidalvel}
\end{figure}

\begin{figure}
\begin{center}
    \leavevmode
    \epsfxsize=7.5cm
    \epsfbox{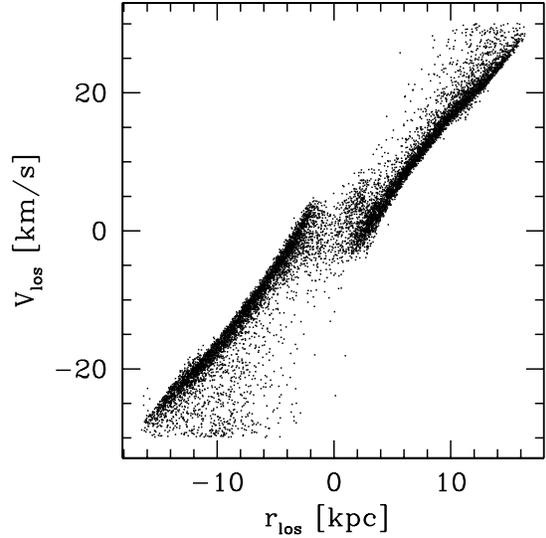}
\end{center}
\caption{Line-of-sight velocities of stars observed along the tidal tails plotted against their position
on the axis parallel to the tails (stars with $r>2.5$ kpc). Only every tenth particle was plotted.}
\label{vel60}
\end{figure}

After the unbound stars are removed from the samples the next most important source of error appears
to originate from the different kinematic properties of the dwarf when observed along different
directions and the measured distribution of the tracer which is one of the components
in dynamical modelling. We have demonstrated that observing the dwarf along the tidal tails we measure
a lower velocity dispersion in comparison with other directions and we see the dwarf as
more extended and more luminous.
The first of these effects decreases the mass, while the second acts in the opposite
direction. However, the first effect is more important so that the mass-to-light ratio
estimated when observing along the tails is {\em underestimated\/} in spite of the strongest
original contamination. Still, our modelling of the velocity dispersion profiles via the Jeans equation
showed that our estimated masses and mass-to-light
ratios agree well (typically within 25 percent) with the true properties of the simulated dwarf.
It should be noted however, that the
mass of the dwarf is best reproduced in the case when the line of sight is perpendicular to the tails.
We also reproduce reasonably well the anisotropy of stellar orbits.

Our results thus show that although the dwarf loses most of its mass due to tidal interaction
and is surrounded by
material which has been stripped from it, it still is a gravitationally bound object which can be
reliably modelled by standard methods based on Jeans equations once interlopers are properly removed.
We emphasize that although we have studied only a single case, our dwarf is a typical product of
the tidal stirring scenario (Mayer et al. 2001, 2002) and was very severely stripped
(it lost 99 percent of its mass) so the result should be even more valid for less stripped satellites.
This conclusion is quite opposite to the tidal scenario as envisioned by Kroupa (1997) and
Klessen \& Kroupa (1998) who claimed that tidal remnants are unbound structures with artificially
inflated velocity dispersions which are useless in determining their true mass-to-light ratios.

Still, this conclusion is far from obvious once we recall (see Fig.~\ref{density})
that some of the material in the tails is still bound to the dwarf because it follows
the dwarf on its orbit but does not trace its potential anymore. We illustrate this in
Fig.~\ref{tidalvel} showing a velocity diagram with 200 stars sampled from the tails alone by
cutting out the main body of
the dwarf up to 2.5 kpc from the centre (we are looking along the tails). The stars unbound to the
dwarf were additionally marked with crosses. The distribution of
velocities is fairly uniform, in particular there are many stars with velocities close to the dwarf
galaxy mean velocity and thus bound to the dwarf but there is no gap between the bound and unbound
stars. The origin of such a velocity distribution can be understood by referring to
Fig.~\ref{vel60}, where line-of-sight velocities of the stellar particles
were plotted against their position on the axis parallel to the tidal tails.
Again, particles with $r<2.5$ kpc from the dwarf's centre were removed. Surprisingly, there seems to be
an almost linear relation between the distance from the
centre of the galaxy and the line-of-sight velocity of the stars. In addition, the tidal tails are
symmetric with respect to the dwarf as they should be due to the symmetric tidal force for small-mass
satellites (Choi, Weinberg \& Katz 2007).

Note, that our procedure of
interloper removal works only on stars with velocities sufficiently discrepant from the galaxy mean.
Given the uniform distribution of velocities in the tails, our procedure for the removal of interlopers
works remarkably well. Still, some of the tidal stars with velocities close to the dwarf's mean remain in
the cleaned samples. We have demonstrated in Fig.~\ref{compileddisp} that these stars do not affect the
velocity dispersion in any significant way and therefore are not really important for dynamical modelling.

We have also studied another possible source of contamination in the dSph kinematic samples,
by the stars from the MW. The importance
of this effect depends however very strongly on the particular location and heliocentric velocity of
the dwarf. Despite the fact that Fornax has greater galactic latitude than Draco
($|b_{\rm For}|=65^\circ$, $|b_{\rm Dra}|=34^\circ$, see {\L}okas et al. 2005 for details on
Draco), contamination from MW stars is expected to be much higher in the
first case, because the mean velocity of Fornax is much closer to the mean velocity of MW stars
in the given direction than in the case of Draco.

We verified the efficiency of our method of interloper removal for samples generated
from one particular configuration of the simulated dwarf contaminated by both tidal tails and
MW stars and found it satisfactory. Still, photometric methods of selection of candidate stars for
spectroscopic observations in dSphs (Majewski et al. 2000a) are to be recommended since they can
save a lot of telescope time.

We applied the methods tested on the simulated samples to the real kinematic data for the Fornax dSph
galaxy from Walker et al. (2006). After removal of interlopers we modelled the velocity dispersion
profile assuming a constant mass-to-light ratio and anisotropy parameter. Contrary to Walker et al.
(2006) we find that the velocity dispersion profile is very well reproduced by a model in which mass
follows light with mass-to-light ratio $M/L =11.3 \; M_{\sun}/L_{\sun}$ and isotropic orbits.
The difference is mainly due to the different
interloper removal scheme applied here. In the case of Fornax, the majority of contaminating stars most
probably comes from the MW and we demonstrated that our interloper removal scheme is successful in
cleaning the sample from them.

\section*{Acknowledgements}

We wish to thank R. Wojtak for the use of his interloper removal code and M. Walker for providing
the photometric data for Fornax stars in electronic form. We are grateful to our referee, M. Wilkinson,
for insightful comments which helped to improve the paper.
We acknowledge discussions with M. Giersz, A. Klypin, S. Majewski and J. Read.
JK and E{\L} are grateful for the hospitality of
Instituto de Astrof{\'\i}sica de Andalucia
in Granada and the Institut d'Astrophysique de Paris where part of this work was done.
SK acknowledges support by the U.S. Department of Energy through a
KIPAC Fellowship at Stanford University and the Stanford Linear
Accelerator Center. The numerical simulations were performed on the
zBox supercomputer at the University of Z\"urich.
We made use of the Besancon Galaxy model available at
http://bison.obs-besancon.fr/modele/.
This research was partially supported by the
Polish Ministry of Science and Higher Education
under grant 1P03D02726, the exchange program of CSIC/PAN and the Jumelage program
Astronomie France Pologne of CNRS/PAN.

\end{document}